\documentclass[manuscript,screen]{acmart}

\AtBeginDocument{%
  }

\setcopyright{acmlicensed}
\copyrightyear{2025}
\acmYear{2025}

\acmISBN{978-1-4503-XXXX-X/18/06}

\usepackage{booktabs}
\usepackage{graphicx}
\usepackage{subcaption}
\usepackage{lscape}

 \usepackage{multirow}
\usepackage{xcolor}

\newcommand{\virgolette}[1]{``#1''}
\newif\ifshowcomments
\showcommentstrue
\ifshowcomments
\newcommand{\mynote}[2]{\fbox{\bfseries\sffamily\scriptsize{#1}}
	{\small\textsf{\emph{#2}}}}
\else
\newcommand{\mynote}[2]{}
\fi

\usepackage{tikz}

\usepackage{changepage}

\newcommand{\mypar}[1]{\par\vspace{0.5em}\noindent\textit{\textbf{#1.}}\hspace{0.1cm}}

\begin{document}

\title[Navigating the Edge-Cloud Continuum: A State-of-Practice Survey]{Navigating the Edge-Cloud Continuum: A State-of-Practice Survey}

\author{Loris Belcastro}
\email{lbelcastro@dimes.unical.it}
\orcid{0000-0001-6324-8108}
\affiliation{%
  \institution{University of Calabria}
  \city{Rende}
  \country{Italy}
}

\author{Fabrizio Marozzo}
\email{fmarozzo@dimes.unical.it}
\orcid{0000-0001-7887-1314}
\affiliation{%
  \institution{University of Calabria}
  \city{Rende}
  \country{Italy}
}

\author{Alessio Orsino}
\email{aorsino@dimes.unical.it}
\orcid{0000-0002-5031-1996}
\affiliation{%
  \institution{University of Calabria}
  \city{Rende}
  \country{Italy}
}

\author{Domenico Talia}
\email{dtalia@dimes.unical.it}
\orcid{0000-0003-1910-9236}
\affiliation{%
  \institution{University of Calabria}
  \city{Rende}
  \country{Italy}
}

\author{Paolo Trunfio}
\email{trunfio@dimes.unical.it}
\orcid{0000-0002-5076-6544}
\affiliation{%
  \institution{University of Calabria}
  \city{Rende}
  \country{Italy}
}


\begin{abstract}
The edge-cloud continuum has emerged as a transformative paradigm that meets the growing demand for low-latency, scalable, end-to-end service delivery by integrating decentralized edge resources with centralized cloud infrastructures. Driven by the exponential growth of IoT-generated data and the need for real-time responsiveness, this continuum features multi-layered architectures. However, its adoption is hindered by infrastructural challenges, fragmented standards, and limited guidance for developers and researchers. Existing surveys rarely tackle practical implementation or recent industrial advances.
This survey closes those gaps from a developer-oriented perspective, introducing a conceptual framework for navigating the edge-cloud continuum. We systematically examine architectural models, performance metrics, and paradigms for computation, communication, and deployment, together with enabling technologies and widely used edge-to-cloud platforms. We also discuss real-world applications in smart cities, healthcare, and Industry 4.0, as well as tools for testing and experimentation. Drawing on academic research and practices of leading cloud providers, this survey serves as a practical guide for developers and a structured reference for researchers, while identifying open challenges and emerging trends that will shape the future of the continuum.
\end{abstract}




\maketitle

\begin{center}
  {\footnotesize\itshape
  Note: This is the author's version of the work. It is posted here for your personal use. Not for redistribution.%
  }
\end{center}

\section{Introduction}
Cloud computing has transformed how modern applications are developed and deployed, offering scalable and cost-efficient processing for a wide range of workloads~\cite{lemos2015web}. Most contemporary services ingest data from diverse sources---such as Internet-of-Things (IoT) sensors, mobile devices, edge nodes, and end users---and rely on centralized cloud resources for aggregation, analysis, and long-term storage~\cite{Hung2019InvestigatingHT}. In recent years, the unprecedented volume of data generated at the network edge has amplified the need for lower end-to-end latency, stronger privacy, and greater scalability~\cite{botta2016integration}. In response, edge–cloud continuum architectures have emerged to bridge the gap between data sources and centralized processing~\cite{moreschini2022cloud}. These multi-layered infrastructures span centralized cloud data centers, decentralized near-edge and far-edge facilities, on-premises environments, and devices located close to data sources. Often referred to as the \textit{cloud continuum}, \textit{cloud-edge continuum}, \textit{IoT-edge-cloud}, or \textit{cloud-to-things continuum}~\cite{kong2022edge}, this paradigm orchestrates computation along the entire Internet-scale path, leveraging each intermediate tier to move processing progressively closer to data producers, while still benefiting from cloud-scale capacity~\cite{ullah2023orchestration}.

However, implementing an edge–cloud infrastructure poses significant challenges. Moving from traditional client–server designs to multi-layer models demands stricter guarantees for privacy, latency, data sovereignty, scalability, and real-time processing~\cite{satyanarayanan2017emergence, 8057318, Ali2021MultiAccessEC}. Addressing this shift requires substantial economic investment to enable the large-scale adoption of edge-cloud architecture. New public cloud data centers must be built across various countries to ensure a consistent low-latency experience for all users,  including the deployment of new on-premises or on-campus micro data centers ~\cite{Eur-cloud-edge-2021}. Additionally, many countries still lack data centers from major public cloud providers, underscoring the need for a network of public edge data centers with a widespread structure to ensure proximity and ultra-low latency. To address this, leading cloud providers, including Amazon, Microsoft, Google, and Alibaba Cloud, are expanding their infrastructure to support edge computing. For example, Amazon Web Services (AWS) has been expanding its network of local data centers, including Local Zones and Edge Locations, across major cities to reduce latency and comply with local data sovereignty regulations. Furthermore, the lack of universal standards remains a significant barrier, hindering interoperability, complicating the seamless integration of distributed applications, and forcing developers to invest additional time and resources in customizing cloud solutions for specific platforms.

Beyond these infrastructural challenges, the edge-cloud continuum has attracted growing attention from both researchers and industry, highlighting its critical role in modern distributed computing. However, the literature presents a fragmented landscape, with varying terminologies, conceptual overlaps, and differing perspectives on architecture and resource management, all contributing to a lack of clarity~\cite{moreschini2022cloud}. Previous surveys in this field have primarily focused on specific aspects such as architecture, resource management, and communication protocols, often overlooking practical considerations and technological solutions crucial for developers building applications across the continuum.
This study, instead, takes a developer-centric approach while maintaining a strong research focus. It bridges key dimensions of the edge-cloud continuum, linking architectural design, models, enabling technologies, deployment platforms, and application domains. By integrating industry-driven developments---covering software, infrastructure, and platforms from major IT companies---with academic contributions in methodologies, algorithms, and tools, this work serves as both a practical guide for developers and a structured reference for researchers navigating the continuum. The key research questions that this survey aims to address are outlined below, each corresponding to the primary topics explored in the subsequent sections.

\begin{itemize}
    \item[\textbf{RQ1}] How should edge-cloud architectures be structured to meet privacy, latency, and scalability requirements?

    \item[\textbf{RQ2}] Which paradigms and models, encompassing deployment, communication, and computation, are most commonly employed in edge-cloud continuum scenarios?

    \item[\textbf{RQ3}] What technologies enable service composition across the edge-cloud continuum, and how do they compare?

    \item[\textbf{RQ4}] What are the major public and private cloud platforms that support service deployment across the edge–cloud continuum, and how do they differ?

    \item[\textbf{RQ5}] What are the key application domains, use cases, and best practices for testing edge-cloud solutions?
    
\end{itemize}

The remainder of the paper is organized as follows. Section~\ref{sec:overview} presents the structure of the survey as a multilayered conceptual framework. Section~\ref{sec:related} analyzes recent surveys on the edge-cloud continuum and discusses the novel aspects of this work. Section~\ref{sec:architecture} presents the multilayered architecture of the edge-cloud continuum. Section~\ref{sec:paradigms} discusses key models and paradigms, while Section~\ref{sec:technologies} introduces the enabling technologies that support service distribution across the continuum. Section~\ref{sec:platforms} examines the main platforms for deploying and managing services along the continuum. Section~\ref{sec:application-domains} discusses tools used for benchmarking these environments and the main application domains. Section~\ref{sec:challenges} identifies and analyzes the open challenges and future research trends in this field, and finally, Section~\ref{sec:conclusion} concludes the paper.

\section{Scope and Contribution}
\label{sec:overview}

\begin{figure}[htb!]
    \centering
    \includegraphics[width=0.7\textwidth]{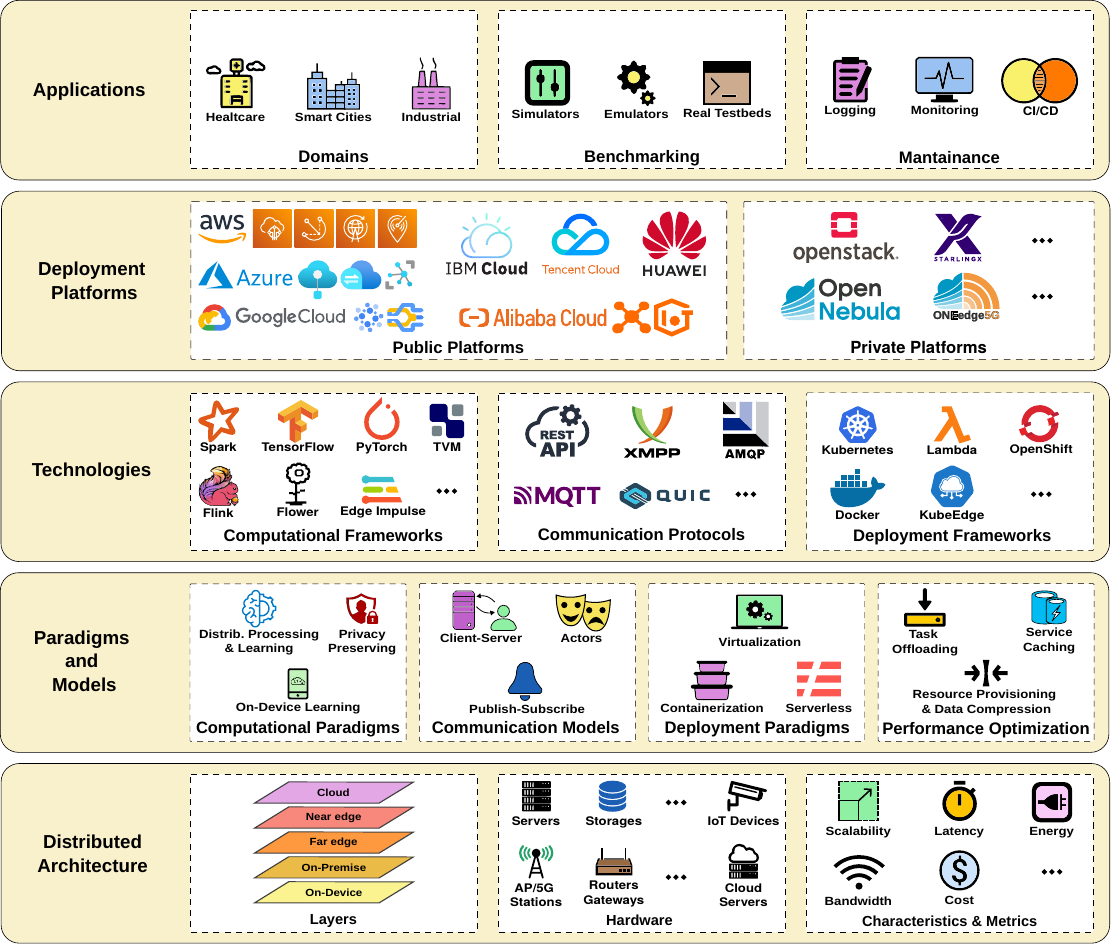}
    \caption{Overview of the proposed conceptual framework for the edge-cloud continuum, illustrating its five key areas of discussion.}
    \label{fig:sections}
\end{figure}

Here we illustrate the vision and contribution of this paper, organized through a conceptual framework designed to guide users in navigating the edge-cloud continuum. As shown in Figure~\ref{fig:sections}, state-of-the-art solutions for the compute continuum are organized into a multilayered framework composed of five distinct areas (\textit{distributed architecture}, \textit{paradigms and models}, \textit{technologies}, \textit{deployment platforms}, and \textit{application domains, use cases and testing tools}), which will be discussed in detail in the subsequent sections of this work.

\begin{figure}[htb!]
    \centering
    \includegraphics[width=1\textwidth]{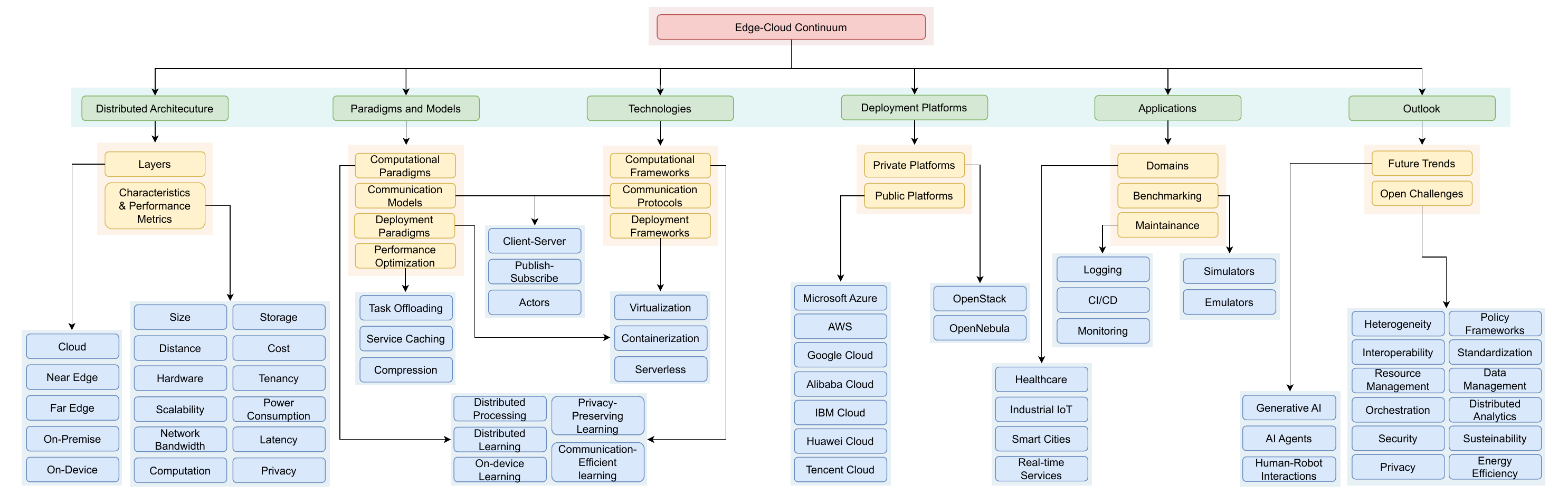}
    \caption{Structure of the survey for navigating the edge–cloud continuum research landscape.}
    \label{fig:sections-overview}
\end{figure}

The first area, discussed in Section~\ref{sec:architecture}, focuses on the \textit{distributed architecture} of edge-cloud systems, covering the main layers such as the cloud, near-edge, far-edge, on-premise, and on-device. It also evaluates critical performance metrics in the design of these systems, such as latency, throughput, scalability, resource utilization, and privacy.
The second area, discussed in Section~\ref{sec:paradigms}, examines the main \textit{paradigms and models} commonly used in edge-cloud systems, from traditional client-server paradigms to more advanced approaches like publisher-subscriber and actor models. Deployment methods, including on-premise setups, virtualization, containers, and serverless deployments are explored. The section also covers computational paradigms such as distributed computing and learning, privacy-preserving learning, and on-device computing. It concludes by discussing performance optimization techniques such as service caching, task offloading, and resource provisioning.
The next area, investigated in Section~\ref{sec:technologies}, explores the \textit{enabling technologies}, including computational frameworks, communication protocols, and orchestration tools. These technologies enable the seamless integration and management of applications across continuum.
The area devoted to \textit{deployment platforms} in Section~\ref{sec:platforms} is another key focus, which compares public platforms like AWS, Azure, Google Cloud, and Alibaba with private solutions such as OpenStack and OpenNebula. Here, we evaluate the capabilities and limitations of these platforms, providing insights into their suitability for various deployment scenarios and their roles in supporting edge-cloud architectures. 
Finally, the last area in Section~\ref{sec:application-domains} explores \textit{application domains and benchmarking tools}. It discusses use cases in key domains such as smart cities, healthcare, industrial IoT, and real-time services. In addition, this section examines the critical phases of \textit{testing} and \textit{maintenance} for edge-cloud continuum applications, highlighting the role of simulators, emulators, testbeds, and CI/CD tools in supporting the development of reliable systems. To help readers navigate the survey more effectively, we present its structure in Figure~\ref{fig:sections-overview}, which systematically organizes the discussion of the edge-cloud continuum along the different dimensions introduced above.
    
\section{Related Surveys}
\label{sec:related}

In recent years, there has been significant interest in the field of edge-cloud continuum systems, leading to a substantial body of literature that surveys various aspects of this domain. Here we review the most relevant surveys, highlighting their contributions to the literature, discussing their different focus, and distinguishing their contents from this work.

Early papers on edge computing explored its architectural design and challenges, including latency reduction, bandwidth efficiency, and security~\cite{shi2016edge, pan_survey}, but did not deeply analyze the distribution of cloud application services across the continuum. Other surveys focused on the role of edge computing in enabling low-latency applications, discussing fundamental concepts and driving forces~\cite{satyanarayanan2017emergence, yu_survey} but lacking a detailed examination of enabling technologies and computational paradigms. Some studies reviewed the state-of-the-art in edge and fog computing, particularly in the context of IoT~\cite{varghese2018next, kong2022edge}, however they missed a holistic perspective on the continuum from the standpoint of cloud application developers.
Further research has addressed the ambiguity surrounding the definition of the edge-cloud continuum, highlighting the diverse interpretations in the literature. Systematic mapping studies have attempted to consolidate these views, providing comprehensive definitions~\cite{moreschini2022cloud}, yet they fall short in offering practical insights for practitioners developing real-world applications in the compute continuum. While comparisons of cloud, fog, and edge computing paradigms have been offered~\cite{mouradian2017comprehensive, aazam2018fog}, discussions often lack a developer-centric focus on service distribution across the continuum. Surveys on fog computing integration with edge and cloud have covered architecture, resource management---focusing on allocation and scheduling---and security~\cite{yousefpour2017fog, hu2017survey, hong2019resource, chiang2023management}, but practical strategies for cloud application adaptation remain underexplored. Evolving telecommunication technologies and their shift toward edge computing to mitigate latency issues have been discussed~\cite{shahzadi2017multi, mao2017survey}.
Lastly, the essential role of edge computing in addressing Internet of Everything (IoE) challenges, particularly in service migration, security, and deployment, has been also articulated~\cite{kong2022edgeeverything, farias2021internet}. 
While there is a rich body of literature surveying various aspects of edge and cloud computing, this survey offers a unique contribution by distinguishing itself from the aforementioned papers in several key aspects:
\begin{itemize}
    \item \textbf{{Developer-centric focus:}} unlike other surveys, this paper adopts the perspective of cloud application developers, addressing their needs and challenges in designing and deploying services across the edge-cloud continuum. This focus is key to understanding how to adapt existing cloud services to decentralized edge environments.

    \item \textbf{{Comprehensive architectural analysis:}} this survey provides a detailed analysis of the architecture of the edge-cloud continuum, including the varying nomenclatures of its layers and the performance metrics associated with each layer. Such a detailed architectural overview is lacking in most existing surveys.

    \item \textbf{{Comparative analysis tools:}} the inclusion of comparative analysis at the end of each section provides a structured approach for developers to evaluate and contrast the different solutions discussed. This method helps developers make informed decisions by offering a clear comparison of pros and cons.

    \item \textbf{{Evaluation of platforms:}} public and open-source cloud platforms are evaluated for their offerings across the continuum, such as computation, storage, and networking. Additionally, the role of \textit{enabling services} in extending cloud-like capabilities to different layers is examined, along with the geographic distribution of different providers and its impact on performance, latency, and accessibility in real-world deployments.

    \item \textbf{{Unified vision:}} 
    this survey provides a broad and unified view of the edge-cloud continuum, integrating various perspectives on available architectures, paradigms, technologies, and platforms, thus offering a cohesive understanding of how cloud and edge solutions can be integrated. 
\end{itemize}

\section{Distributed Architecture and Metrics}
\label{sec:architecture}

This section provides an in-depth examination of the core concepts, architectural layers, performance metrics, and key characteristics of edge-cloud computing architectures. A thorough understanding of these layers and their characteristics is essential for designing efficient, scalable, and secure systems, as highlighted in previous research~\cite{aslanpour2020performance}.

\subsection{Architectural Layers}
The edge-cloud continuum is organized as a hierarchical architecture comprising multiple layers, each serving distinct purposes and integrating complementary functionalities to support efficient computing and data management~\cite{gkonis2023survey, balouek2019towards}. In a classic edge-cloud continuum system, these layers are strategically arranged to optimize data storage, processing, and analysis while ensuring low-latency communication and effective workload distribution~\cite{rosendo2022distributed}.
From a bottom-up perspective, the foundational layer is the \textit{device layer}, which includes a diverse array of edge devices such as smartphones, GPS units, onboard cameras, IoT sensors, wearable devices, and connected vehicles. These devices generate raw data and may perform preliminary tasks, such as filtering, aggregation, compression, and localized decision-making, to reduce latency and network overhead before transmitting information to nearby edge servers~\cite{gkonis2023survey}.
The \textit{edge layer} comprises hardware components like gateways, micro data centers, edge routers, and local processing nodes. These elements collect data from the device layer and execute time-sensitive processing tasks near the data source~\cite{khalyeyev2022towards}. In some architectures, a \textit{fog layer} acts as an intermediary between the edge and cloud layers. This layer offloads heavy computational tasks and mitigates latency by enhancing resource allocation and overall system efficiency.
At the top of the hierarchy, the \textit{cloud layer} offers scalable computing and storage resources for complex tasks beyond the capabilities of edge and fog layers, such as large-scale data analytics, advanced machine learning, and long-term historical data storage~\cite{gkonis2023survey}. This hierarchical structure facilitates seamless data flow and efficient resource utilization across the continuum, enabling optimized performance and enhanced computational capabilities in diverse applications~\cite{ khalyeyev2022towards}. 

\begin{figure}[htb!]
    \centering
    \includegraphics[width=0.7\textwidth]{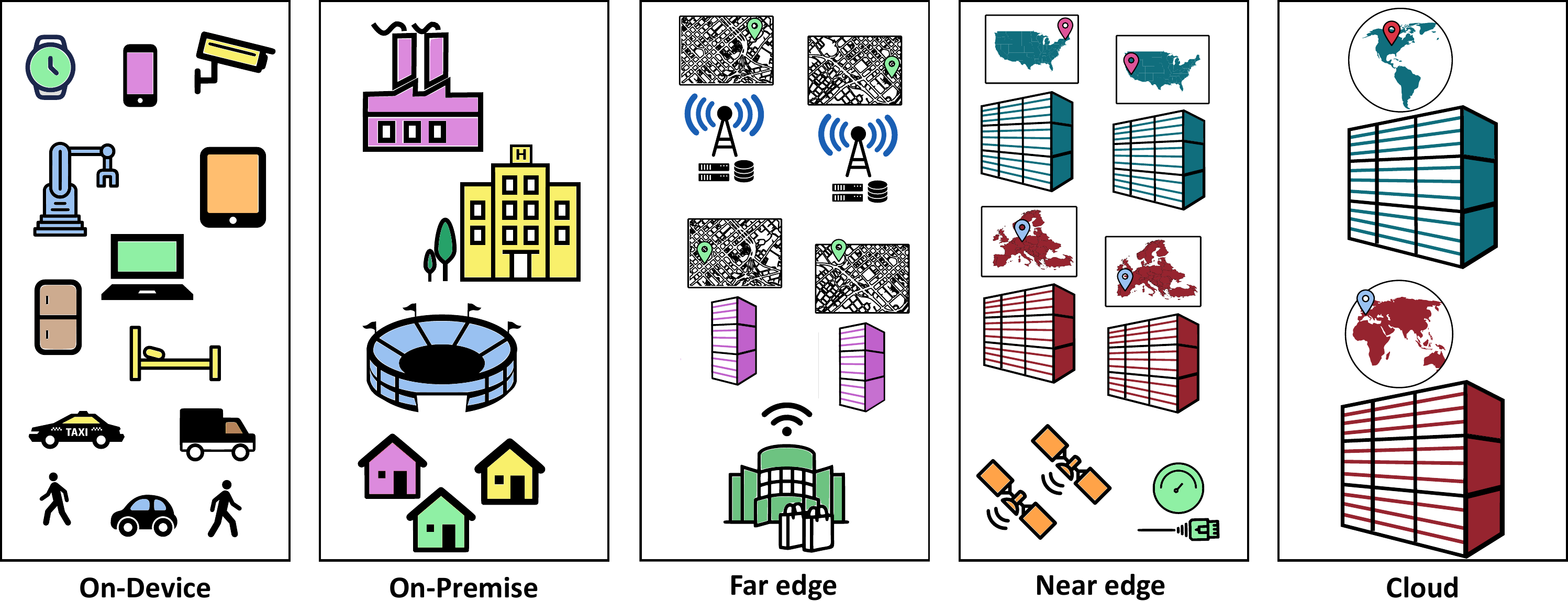}
    \caption{Edge-cloud continuum layers.}
    \label{fig:solu-arch}
\end{figure}

Recently, a novel edge-cloud architecture, shown in Figure \ref{fig:solu-arch}, has been introduced to address emerging requirements from industry and governments~\cite{Eur-cloud-edge-2021}, including:
$i)$ the growing need to maintain autonomy and sovereignty over edge and cloud technologies;
$ii)$ the increasing electrical power demand driven by the widespread adoption of cloud computing; and
$iii)$ the rising need for on-premise micro data centers to ensure extreme privacy and low latency.

This alternative architecture introduces new intermediate layers within the edge-cloud continuum, adopting a cloud-centric perspective that classifies them based on their proximity to the cloud rather than to end-user devices:

\begin{itemize}
\item \textbf{\textit{Cloud}}: this layer serves as the central hub for large-scale data processing and storage. It includes public data centers provided by major cloud service providers such as Amazon Web Services (AWS), Microsoft Azure, and Google Cloud Platform. Applications run on remote servers managed by these providers, which also offer asset management, security, and monitoring services. Hardware and software resources are shared in a multi-tenant environment across different organizations and users.

\item \textbf{\textit{Near edge}}: positioned closer to the cloud than the traditional edge, this layer facilitates faster data processing before reaching core cloud infrastructure. It consists of mini data centers, central offices, or regional cloud nodes that connect near-edge resources to the cloud. These facilities, though smaller than those in the cloud layer, still operate in a multi-tenancy mode and can be located several hundred kilometers away from devices.

\item \textbf{\textit{Far edge}}: this layer brings computing resources even closer to end-user environments, ensuring faster data aggregation and preliminary processing. It includes nodes deployed at mobile phone towers, near large shopping malls, or adjacent to industrial sites. These small-scale public data centers and specialized networking equipment handle localized computing tasks before forwarding processed data to higher-tier infrastructure.

\item \textbf{\textit{On-premise}}: this layer consists of data processing nodes operating within local or end-user facilities, such as farms, stadiums, or manufacturing plants. It balances proximity and autonomy by deploying private edge nodes very close to devices, ensuring fast, stable, and real-time connectivity for business or user systems.

\item \textbf{\textit{On-device}}: the closest layer to end users, it comprises IoT devices that collect and process data locally, reducing dependency on external infrastructure and enabling real-time decision-making.
\end{itemize}

At the near edge, it is worth highlighting the role of \textit{on-ramps} in optimizing connectivity between edge environments and the cloud. These on-ramps typically consist of a combination of physical facilities (e.g., carrier-neutral data centers and internet exchange points), dedicated high-performance network links (e.g., private interconnects), and logical services (e.g., traffic routing optimization). \textit{Satellites} act as special-purpose on-ramps, enabling cloud access where terrestrial infrastructure is unavailable, such as in remote areas. 

\subsection{Characteristics and Performance Metrics}

Understanding the characteristics of edge-cloud systems is crucial for optimizing resource allocation, minimizing latency, ensuring data privacy, and managing operational costs. In this section, we examine key attributes and performance metrics across the different layers of the compute continuum, from on-device to cloud layers~\cite{aslanpoura2020internet, charyyev2020latency, maheshwari2018scalability, cao2020overview, Kimovski2021CloudFO, omran2024edge}. Table~\ref{tab:architecuture-comparison} offers a comprehensive comparison of the different layers of an edge-cloud computing architecture based on these characteristics, providing specific data or ranges of values for each one of them.

\begin{table}[h!]
\fontsize{7pt}{7pt}\normalfont{
\centering
\setlength{\tabcolsep}{2.5pt} 
\begin{tabular}{@{}llllll@{}}
\toprule
\textbf{Metric/Layer} &
  \textbf{On-device} &
  \textbf{On-premise} &
  \textbf{Far edge} &
 \textbf{Near edge} &
  \textbf{Cloud} \\ \midrule
\begin{tabular}[c]{@{}l@{}}\textbf{Size}\end{tabular} &
  1M &
  100k &
  100-1000 &
  10 &
  \textless{}10 \\
\addlinespace[0.7ex]

\textbf{Scalability} &
  Very High &
  High &
  Medium &
  Low &
  Very Low \\
\addlinespace[0.7ex]

\textbf{Hardware} &
  \begin{tabular}[c]{@{}l@{}}IoT devices, cameras,\\ sensors, smartphones\end{tabular} &
  \begin{tabular}[c]{@{}l@{}}Small servers installed\\ in the close poximity of \\ datasources (e.g., gateway,\\stadiums, base stations, \\street lights, road side units)\end{tabular} &
  \begin{tabular}[c]{@{}l@{}}Physical containers or aggreg.\\ nodes to cover local hotspots\\ (e.g., shopping, business \\or touristic areas)\end{tabular} &
  \begin{tabular}[c]{@{}l@{}}Mini datacenters, \\e.g., regional or in-country \\datacenters, to cover\\ specific wide or \\crowded areas \end{tabular}&
  \begin{tabular}[c]{@{}l@{}}Datacenters to\\ cover worldwide \\areas\end{tabular}\\
\addlinespace[0.7ex]

\textbf{Distance} &
  0 &
  \textless 1km &
  1-100km &
  100-1000km &
  \textgreater 1000 km \\
\addlinespace[0.7ex]

\textbf{Latency} &
  \textless 1ms &
  1ms &
  2-5ms &
  10-20ms &
  \textgreater 20ms \\
\addlinespace[0.7ex]

\begin{tabular}[c]{@{}l@{}}\textbf{Network} \\ \textbf{bandwidth}\end{tabular} &
  \begin{tabular}[c]{@{}l@{}}Very low \\ (KBps-MBps)\end{tabular} &
  \begin{tabular}[c]{@{}l@{}}Moderate \\ (MBps)\end{tabular} &
  \begin{tabular}[c]{@{}l@{}}Moderate to High \\ (MBps-Gbps)\end{tabular} &
  \begin{tabular}[c]{@{}l@{}}High \\ (Gbps)\end{tabular} &
  \begin{tabular}[c]{@{}l@{}}Very High \\ (Gbps-Tbps)\end{tabular} \\
\addlinespace[0.7ex]

\begin{tabular}[c]{@{}l@{}}\textbf{Computation} \\ \textbf{capabilities}\end{tabular} &
  Very Low &
  Low &
  Medium &
  High &
  Very High \\
\addlinespace[0.7ex]

\begin{tabular}[c]{@{}l@{}}\textbf{Storage}\\ \textbf{capabilities}\end{tabular} &
  Very Limited &
  Limited &
  \begin{tabular}[c]{@{}l@{}}Limited (cache only), \\not suitable for \\persistent storage\end{tabular} &
  Medium &
  \begin{tabular}[c]{@{}l@{}}Very High \\(Unlimited)\end{tabular} \\
\addlinespace[0.7ex]

\textbf{Cost} &
  $\sim$5k€ &
  $\sim$100k€ &
  $\sim$0.5M€ &
  $\sim$10M€ &
  $\sim$2.9B€ \\
\addlinespace[0.7ex]

\begin{tabular}[c]{@{}l@{}}\textbf{Power} \\ \textbf{consumption}\end{tabular} &
  \textless 1kW &
  20-50 kW &
  30-100kW &
  0.5-1MW &
  5-100MW \\
\addlinespace[0.7ex]

\begin{tabular}[c]{@{}l@{}}\textbf{Tenancy}\end{tabular} &
  Dedicated &
  Dedicated &
  Multi-tenancy &
  Multi-tenancy &
  Multi-tenancy \\
\addlinespace[0.7ex]

\textbf{Privacy} &
  Very high &
  High &
  High-Medium &
  Medium-Low &
  Low \\
\addlinespace[0.7ex]

  \bottomrule
\end{tabular}
\caption{Comparison of different layers in the edge-cloud continuum based on key characteristics.}
\label{tab:architecuture-comparison}
}
\end{table}

\mypar{Size} This feature indicates the number of devices involved at each layer of the edge-cloud continuum. The on-device layer typically involves a vast number of endpoints, potentially reaching millions, while as we move toward the cloud fewer devices are utilized. For example, the cloud layer operates with few globally distributed data centers.

\mypar{Scalability} It indicates the capability of each layer to support the addition of new devices, users, and services. For instance, the on-device layer exhibits very high scalability, as it can easily accommodate new devices. On-premise facilities also show high scalability, while far-edge deployments tend to offer moderate scalability. Moving to the cloud, near-edge systems show low scalability, whereas cloud infrastructure has the lowest scalability in terms of adding physical nodes, though it excels in vertical scalability through resource expansion.

\mypar{Hardware} This dimension pertains to the hardware components commonly used at each layer. At the device layer, infrastructure encompasses a diverse array of endpoints, including IoT devices, cameras, sensors, smartphones, and wearables. On-premise edge infrastructure consists of small servers strategically positioned near data sources. Far-edge deployments utilize physical containers or aggregation nodes to cover local hotspots, such as shopping, business, or tourist areas. In near-edge environments, mini-datacenters, such as in-country datacenters, are typically deployed, whereas cloud infrastructure relies on large-scale data centers distributed worldwide.

\mypar{Distance} It denotes the geographical span covered by each layer. For instance, on-device processing occurs directly at the data source. On-premise edge computing extends up to a kilometer from the data sources, while far-edge deployments cover distances ranging from 1 to 100 kilometers. Near-edge systems span distances of 100 to 1000 kilometers, whereas cloud infrastructure operates on a global scale, covering distances exceeding 1000 kilometers~\cite{Eur-cloud-edge-2021}.

\mypar{Latency} This feature measures the time delay incurred during data transmission and processing within each layer of the edge-cloud architecture. On-device processing achieves ultra-low latency due to its immediate proximity to data sources. As the distance between the data source and the processing unit increases, moving from on-premise edge to far-edge, near-edge, and finally to the cloud, latency progressively rises. This increase is primarily due to longer transmission distances and additional network hops, which introduce delays in data propagation and processing.

\mypar{Network bandwidth} This facet measures the data transfer rate and capacity available at each layer. On-device processing relies on minimal bandwidth, primarily limited to local communication between sensors, actuators, or embedded systems. On-premise and far-edge layers typically provide moderate bandwidth for localized data exchanges and edge-to-cloud communication. Near-edge systems, often connected via high-speed networks, offer higher bandwidth to support regional data aggregation and processing. Finally, cloud infrastructure relies on extremely high bandwidth, supported by robust backbone networks, to manage large-scale data flows and ensure global accessibility.

\mypar{Computation capabilities} This characteristic refers to the processing capabilities available at each layer of the edge-cloud system. On-device processing exhibits very low computation power due to limited hardware resources, suitable for basic data collection and preprocessing tasks. On-premise edge computing offers low to medium computation power, sufficient for common data analysis and decision-making. Far-edge and near-edge layers provide medium to high computation power, supporting advanced analytics. In contrast, cloud infrastructures offer very high computation power, enabling large-scale big data processing and AI-driven analysis.

\mypar{Storage capabilities} This dimension assesses the capacity and persistence of data storage at each layer of the edge-cloud architecture. The on-device layer offers very limited storage capacity, typically suited for temporary data buffering or caching. On-premise and far-edge layers provide limited storage for local or transient data needs. Near-edge facilities offer medium storage capabilities, sufficient for regional caching and data persistence. Instead, cloud infrastructure provides virtually unlimited storage capacity, supporting large-scale data warehousing and archival and ensuring long-term data persistence.

\mypar{Cost} This aspect measures the financial expenditure associated with each layer. Costs vary significantly across layers, with on-device processing being relatively inexpensive. However, as the system scales toward cloud infrastructures, costs increase significantly. For example, modern cloud deployments can average around 2.9 billion euros per deployment, mainly due to the establishment and maintenance of large-scale data centers~\cite{Eur-cloud-edge-2021}.

\mypar{Power consumption} It measures the energy usage of each layer. On-device processing consumes less than 1 kW per hour, whereas on-premise edge computing consumes between 20 and 50 kW per hour. Far-edge deployments consume between 30 and 100 kW per hour, while near-edge systems consume between 0.5 and 1 MW per hour. Cloud infrastructures are the most energy-intensive, with consumption ranging from 5 to 100 MW per hour, depending on the scale and type of datacenter~\cite{Eur-cloud-edge-2021}.

\mypar{Tenancy} This feature refers to the degree of resource sharing and isolation. The on-device and on-premise layers typically maintain dedicated infrastructure, ensuring exclusive resource access for individual applications. In contrast, the far-edge, near-edge, and cloud layers often adopt a multi-tenancy approach, enabling shared resource utilization among multiple applications. This reflects also the service distribution model, which defines how resources and services are provisioned and accessed. The on-device and on-premise layers typically employ private service distribution models, while the far-edge, near-edge, and cloud layers adopt public service distribution models, supporting multi-tenancy.

\mypar{Privacy}
Finally, privacy refers to the protection of sensitive data and user information. Local data processing and dedicated infrastructures at the on-device and on-premise layers generally yield higher levels of privacy. Conversely, far-edge, near-edge, and cloud layers might present varying degrees of privacy risk, especially when data is transmitted and stored across multiple jurisdictions. 

\section{Paradigms and Models}
\label{sec:paradigms}

The successful development, deployment, and execution of distributed applications in the edge-cloud continuum rely on leveraging appropriate models and paradigms that accommodate resource availability, performance goals, and application requirements. These models abstract the complexities of the underlying infrastructure, enabling seamless distribution of computation and data across edge devices, intermediate nodes, and cloud platforms.

In the following sections, we explore the foundational paradigms that drive this continuum. We begin by examining computational models that enable distributed processing, focusing specifically on AI-based approaches to intelligent data handling, from collaborative, cloud-assisted training to on-device analytics. Next, we discuss communication models, including client-server, publish-subscribe, and actor-based paradigms, which facilitate data exchange and coordination across heterogeneous networks. We then transition to deployment paradigms such as virtualization, containerization, and serverless computing, which offer the scalability required for dynamic application environments. Complementing these discussions, we also examine performance optimization strategies, such as task offloading and service caching, which help mitigate latency, reduce network congestion, and enhance system responsiveness.

\subsection{Computational Paradigms}
In the edge-cloud continuum, data analytics tasks are distributed across edge and cloud environments to optimize performance and efficiency. At the edge, preprocessing tasks such as filtering, aggregation, and basic inference reduce data volume before transmission to the cloud~\cite{Ghosh2019DeepLE}. The cloud, instead, handles complex tasks like large-scale analytics. Particularly, artificial intelligence and machine learning (ML) have become key tools of modern data analytics, enabling systems to learn from data and make intelligent decisions. In the context of the edge-cloud continuum, AI-based analytics leverages both edge and cloud resources to optimize performance and efficiency~\cite{s22072665}. Computation can occur in a collaborative manner, where multiple devices at different levels of the compute continuum cooperate to train a global model, also exploiting recent paradigms focused on privacy-preserving and communication-efficient learning~\cite{prigent2024enabling, pfeiffer2023federated}, such as federated and split learning~\cite{thapa2021advancements}. In contrast to collaborative learning, recent advances in hardware and model optimization have led to the development of the on-device computing paradigm, where machine learning models are trained directly on individual devices to minimize the need of data transfer, enhance privacy, and allow for real-time model updates~\cite{Dhar2021ASO}.

\mypar{Distributed Processing}
Distributed processing supports generic data analytics tasks in geographically distributed and resource-diverse environments through two main paradigms: batch and stream processing. \textit{Batch processing} handles large, static data sets collected over time, making it ideal for throughput-intensive tasks with low latency. Common edge-cloud uses include complex data transformations, historical analysis, periodic reporting (e.g., hourly, daily), and data warehousing, typically centralized in cloud infrastructures but sometimes initially preprocessed at edge gateways~\cite{8822056}. 
On the other hand, \textit{stream processing} handles continuous, real-time data streams, focusing on low-latency responses. It is essential at the edge for tasks such as immediate sensor data filtering, anomaly detection, data quality assessments, and alerting~\cite{8014357}. Processed data or event notifications are often sent to the cloud for further aggregation, correlation analysis, or persistent storage. Hybrid approaches that combine batch and stream processing are commonly used to analyze historical data while responding quickly to real-time events. These tasks are typically orchestrated using directed acyclic graphs (DAGs), facilitating task dependency management, scheduling, resource allocation, and fault tolerance in distributed environments~\cite{DIASDEASSUNCAO20181,rosendo2022distributed}.

\mypar{Distributed Learning}
Distributed machine learning algorithms can be implemented using two different approaches: distributing the data or distributing the model~\cite{kraska2013mlbase}. In the data-parallel approach, data is partitioned across clients, which all execute the same algorithm on different partitions of the data. The different models obtained by training the algorithm on the various partitions are then aggregated by the server. In the model-parallel approach, instead, the same data is processed by clients, which execute different partitions of the same model, and the final model is therefore generated by the aggregation of all parts. This approach can be applied to all those machine learning algorithms in which model parameters can be partitioned, such as neural networks. Another approach is based on ensemble learning~\cite{verbraeken2020survey}, in which several instances of the same model are trained and used for inference, aggregating the outputs coming from each model. In all of these approaches, worker nodes can be arranged in either a centralized architecture, also known as parameter server architecture~\cite{li2013parameter}, or in a decentralized one. The parameter server architecture consists of one or more servers and several workers, and the learning process is performed iteratively by updating and synchronizing model parameters with central servers~\cite{Li2014ScalingDM}. Instead, in the decentralized setting, each worker communicates with its neighbors and the model is aggregated without a central coordination. 

\mypar{Communication-Efficient and Privacy-Preserving Learning}
Federated learning (FL) is a collaborative learning paradigm that enables multiple clients to train a model while keeping their data decentralized, in contrast to traditional machine learning where data is centrally stored or transmitted to remote cloud servers~\cite{konevcny2016federated}. This approach addresses concerns such as data privacy and data transfer minimization, making it particularly relevant in privacy-sensitive fields such as healthcare~\cite{sheller2020federated} and finance~\cite{long2020federated}. The core idea is to train a model on multiple local datasets across distributed clients without sharing the actual data, by exchanging only the parameters (e.g., weights and biases of a neural network). While traditional distributed learning typically assumes independent and identically distributed datasets of similar size, involving homogeneous nodes with powerful computational capabilities such as data centers connected by fast networks, federated learning focuses on training across heterogeneous clients and data of varying distributions. Moreover, clients in FL systems are often less reliable and may experience more frequent failures due to their reliance on less robust communication protocols and battery-powered systems. Split Learning (SL) is another collaborative learning paradigm that allows training models without necessitating data sharing~\cite{thapa2021advancements}. Unlike federated learning, SL partitions the model into segments, which are trained on different clients, and only the weights of the final layer from each segment are transmitted to the subsequent client, ensuring model learning while preserving data privacy, also making this paradigm more suitable for resource-constrained devices. ~\cite{Singh2019DetailedCO}.

\mypar{On-device Machine Learning}
Deploying machine learning at the edge enables low-latency training and inference directly on data sources, benefiting various real-world applications~\cite{merenda2020edge}.Moreover, on-device learning allows models to adapt to user behavior and preferences in real time, enabling highly personalized experiences~\cite{Zhou2021OnDeviceLS}. For example, in healthcare, wearable devices can analyze personal health data to provide timely and tailored health advice~\cite{incel2023device}. However, the limited computational and energy power, the heterogeneity in hardware and technologies, and security issues of IoT edge devices pose a great challenge in performing learning tasks on such devices~\cite{merenda2020edge}. The conventional approach involves training large models using high-performance computing (HPC) clusters in the cloud~\cite{shuvo2022efficient} and compressing them using techniques like knowledge distillation~\cite{hinton2015distilling, cantini2024xai}, pruning~\cite{liu2018rethinking}, and quantization~\cite{zhang2018lq}. Instead, on-device training is much less common due to computational
limitations~\cite{kukreja2019training}. To overcome these challenges, meta-learning paradigms allow models to quickly adapt to new tasks with minimal data and computation~\cite{zhu2024device}. Furthermore, to optimize models for ultra-low-power devices, such as microcontrollers (MCUs), Tiny machine learning (TinyML) leverages techniques such as neural architecture search (NAS)~\cite{Ray2021ARO} and incremental and continual learning~\cite{10.1145/3417313.3429378}, which help update models over time while minimizing memory and compute requirements. 

\subsection{Communication Models} 
Communication models define how components in a distributed system interact, shaping the architecture and behavior of devices in edge and cloud environments by regulating data exchange and coordination. In the following, the most important communication models, client-server, publish-subscribe, and actor-based models are presented.

\mypar{Client-Server} The client-server model remains a foundational paradigm within the edge-cloud continuum, where edge devices act as clients that request services or resources, while cloud servers process these requests and deliver responses. The client-server pattern is well-suited for applications requiring centralized control and resource-intensive computations, as it allows clients to offload heavy tasks to more capable servers~\cite{jansen2023continuum, al2024computing}. In edge-cloud environments, this model is often extended to include intermediate far- and near-edge layers, which act as local servers to handle latency-sensitive tasks closer to the data source. This multi-tier adaptation of the client-server model helps bridge the gap between centralized cloud services and resource-constrained edge devices, ensuring efficient data processing and service delivery across the continuum~\cite{baresi2019unified}.

\mypar{Publish-Subscribe} The publish-subscribe model facilitates asynchronous communication between data producers and consumers, by decoupling publishers and subscribers and enabling scalability and flexibility in managing dynamic workloads such as those of distributed systems and IoT networks. Hierarchical publish-subscribe models reduce latency and optimize resource usage by clustering brokers close to edge devices~\cite{pham2019efficient, pham2021efficient}. Additionally, multi-tier computational models that integrate publish-subscribe systems, as explored in~\cite{veeramanikandan2019publish}, demonstrate their effectiveness in supporting large-scale IoT applications. This paradigm is particularly suitable for real-time applications requiring efficient data dissemination across geographically distributed nodes, enabling scalable communication.

\mypar{Actors} The actor model provides a distributed, event-driven way for building highly concurrent systems. In this model, actors encapsulate state and behavior and communicate asynchronously through message passing. This decentralized approach enhances scalability and fault tolerance, particularly in systems with complex, hierarchical workloads as the edge-cloud continuum. Actor-based frameworks are frequently used to implement distributed fog computing applications, ensuring efficient task delegation and coordination~\cite{srirama2021akka}. Furthermore, the actor model can adapt to workload changes, making it ideal for edge IoT applications that require high reactivity and autonomy~\cite{zhang2021collaboration}.

\subsection{Deployment Paradigms}
Deployment paradigms provide high-level design strategies for abstracting infrastructure and organizing system components and their interactions. These patterns support the flexible deployment of workloads across different layers of the compute continuum, using approaches like virtual machines, containers, and serverless functions. 
A key enabler of these paradigms is the microservice software development approach, which structures applications as a collection of small, loosely coupled, and independently deployable services. In the edge–cloud continuum, microservices enable fine-grained task management and adaptability to resource constraints across layers. By employing decentralized orchestration mechanisms, microservices can efficiently manage heterogeneous resources while maintaining service continuity~\cite{kiss2024decentralised}. Optimized microservice placement strategies ensure that latency-sensitive components are deployed on edge nodes, while computationally intensive tasks are offloaded to the cloud~\cite{mota2024optimizing}.

\mypar{Virtualization}
Virtualization in the edge-cloud continuum enables the seamless deployment and management of applications from multiple users on a shared infrastructure, ensuring key features such as isolation, fault tolerance, and resource efficiency~\cite{tao2019survey, Ramalho2016VirtualizationAT}. Virtual Machines (VMs) provide an abstraction layer over the underlying hardware, allowing applications to run consistently across heterogeneous environments, regardless of the physical device’s architecture. This capability supports scalability and interoperability, ensuring efficient resource sharing and monitoring. A recent advancement in virtualization is the emergence of MicroVMs, which offer a lightweight approach that balances the isolation and security of traditional VMs with the efficiency of containers~\cite{Lee2023MicroVMOE}. Unlike full-fledged VMs, which require a dedicated operating system instance, MicroVMs include only the essential components needed for a specific workload, reducing startup time, memory footprint, and CPU overhead. MicroVMs enable secure containerized applications and lightweight virtualized workloads with strong isolation and minimal overhead. 

\mypar{Containerization}
Containers represent a lightweight form of virtualization that is particularly well-suited for edge computing. Unlike traditional VMs, containers share the host system’s kernel, eliminating the need forService a full operating system instance and reducing resource overhead. This results in faster startup times, lower memory consumption, and improved deployment~\cite{carpio2020engineering}. Containers enhance portability and scalability, ensuring that applications run consistently across diverse hardware and software configurations. This flexibility is particularly beneficial in edge computing, where applications must dynamically scale and relocate based on network conditions, resource availability, and workload distribution. By bringing computation closer to data sources, containers reduce latency and optimize performance~\cite{Morabito2017VirtualizationOI}. However, in edge environments with limited resources, efficient resource allocation strategies are essential to maintain predictable performance and avoid resource contention.  
Both VMs and containers can exploit consolidation techniques to improve resource utilization and energy efficiency~\cite{Gholipour2020ANE}. Since containers often run within VMs, optimizing their joint placement can reduce resource fragmentation and improve utilization while minimizing migration overhead. To this end, extensive studies have explored adaptive consolidation strategies that balance workload distribution, energy consumption, and performance, while minimizing Service Level Agreement (SLA) violations and latency~\cite{ding2020adaptive, Farahnakian2019EnergyAwareVC}.

\mypar{Serverless deployment} 
Serverless deployment is a cloud execution model that simplifies application distribution by allowing developers to write and deploy functions (or micro-tasks) without worrying about the underlying infrastructure, as it abstracts infrastructure management~\cite{Raith2023ServerlessEC}. In the edge-cloud continuum, serverless functions are particularly valuable for handling event-driven workloads~\cite{risco2021serverless} and scaling resources dynamically~\cite{Raith2023ServerlessEC}. These functions can migrate across layers, adapting to real-time conditions such as network latency and resource availability~\cite{russo2024framework}. Serverless computing is also being integrated with collaborative learning paradigms to extend its utility to IoT applications, allowing seamless interaction between cloud and edge layers~\cite{loconte2024expanding}. Moreover, emerging paradigms like osmotic computing combine serverless workflows with security-enhanced architectures for critical edge-cloud applications~\cite{morabito2023secure}.
A key enabler of serverless computing is the broader \virgolette{as-a-Service paradigm}, which has evolved from traditional cloud service models such as Infrastructure-as-a-Service (IaaS) and Platform-as-a-Service (PaaS) to more sophisticated and granular approaches. In this context, \textit{Function-as-a-Service (FaaS)} emerged as an effective model for supporting the execution of parallel and geographically distributed applications in the edge-cloud continuum~\cite{benchfaas,10.1145/3437378.3444367}. FaaS enables users to deploy and run self-contained computational functions in a fully serverless manner~\cite{schleier2021serverless}, eliminating the complexities of provisioning infrastructure and software.

\subsection{Performance Optimization}
Optimizing performance in edge-cloud environments is essential for maximizing resource efficiency and minimizing latency. Techniques such as task offloading, service caching, and data compression are pivotal in tackling the challenges inherent to the edge-cloud continuum.

\mypar{Task offloading} Task offloading aims at optimizing resource utilization in edge-cloud systems, enhancing application performance and mitigating energy consumption within edge device~\cite{islam2021survey, ullah2023optimizing}. Indeed, offloading computationally intensive tasks from devices to far-edge, near-edge, and cloud servers with higher processing power can significantly reduce latency~\cite{liu2024fast}. Moreover, since edge devices often have limited battery life, it can save energy and extend their operational lifespan~\cite{li2022task}. Last, balancing the workload across edge and cloud resources is essential for maximizing application throughput and scalability~\cite{saeik2021task}. Task offloading strategies in edge-cloud environments can be broadly categorized into \textit{static offloading}, where decisions are made beforehand based on predetermined rules or heuristics~\cite{almutairi2021novel}, and \textit{dynamic offloading}, where decisions are made in real-time, considering factors like network latency, device capabilities, application characteristics, and security constraints. 
Another way to categorize task offloading techniques is based on the method used for decision-making. One common approach is \textit{heuristic-based} methods, including genetic algorithms, ant colony optimization, and simulated annealing~\cite{guo2021hagp,nguyen2023dependency,abbas2021meta}. While heuristics offer approximated offloading solutions, they may not adapt well to dynamic environments. To address this limitation, \textit{AI-driven} approaches have gained prominence, leveraging machine learning to optimize offloading decisions based on historical data and real-time analytics. Specifically, reinforcement learning techniques have been widely adopted due to their ability to learn and adapt dynamically~\cite{ullah2023optimizing,tang2020deep,qu2021dmro,fang2022ai}. However, AI models typically assume a centralized decision-making framework, which may not always be suitable. In such scenarios, \textit{game theory-based} approaches become particularly relevant to modeling strategic interactions between edge devices and the cloud, by enabling negotiation and cooperation in multi-agent environments where decentralized decision-making is required~\cite{su2021game,chen2022qoe,xu2022game,teng2022game}. 
To further enhance trust and security in task offloading, blockchain-based approaches have emerged as a complementary solution, ensuring data integrity and transparency between edge and cloud environments~\cite{li2024blockchain,fan2023blockchain,yao2022blockchain,samy2022secure}.

\mypar{Service caching} Service caching is another key strategy for optimizing resource management in edge-cloud systems. By storing frequently accessed services closer to end-users, such as in Mobile Edge Computing (MEC)-enabled Base Stations (BS), latency and network congestion are reduced~\cite{ni2020security}. Various paradigms have emerged to address key decisions regarding what services to cache, where to place them, and when to update or evict them. Dynamic adaptation leverages optimization techniques to balance multiple objectives and continuously adjust caching decisions in response to fluctuating service demand, network conditions, and resource availability~\cite{xu2018joint}. However, this real-time process introduces significant computational complexity. Instead, approaches based on predictive models aim to proactively anticipate service demand, determining which services are likely to be needed in the near future. By leveraging historical data and machine learning models, predictive caching minimizes unnecessary cache evictions and preemptively places services~\cite{xie2018dynamic}, reducing response times and network overhead. 
To further optimize service caching, collaborative caching strategies enable cooperative decision-making among multiple caching entities through joint coordination between edge devices and cloud servers~\cite{huang2021enabling}, enhancing system efficiency by sharing information between nodes.

\mypar{Resource provisioning and data compression} 
Another paradigm in performance optimization in edge-cloud systems is adaptive resource provisioning, where resources are allocated dynamically based on real-time workload demands~\cite{duc2019machine}. By continuously monitoring and adjusting resource allocation, this approach significantly enhances latency and utilization costs while preventing performance bottlenecks caused by hardware limitations~\cite{shakarami2022resource}.  
A further optimization can be performed at the data level. One method to reduce cloud bandwidth consumption is to compress raw data at the edge before uploading it to the cloud~\cite{Wang2020JointOO}. Generally, lossy compression reduces data size at the cost of losing details, which can severely impact the quality of analytics based on the compressed data. Therefore, it is critical to develop data compression methods that minimize communication costs while preserving computational accuracy~\cite{dong2020cdc}.

\section{Enabling Technologies}
\label{sec:technologies}

The edge-cloud continuum involves many technologies spanning various disciplines, including telecommunications, industrial automation, and information technology (IT). While this survey acknowledges the importance of these diverse technological areas, this section analyzes the enabling technologies for the edge-cloud continuum from an IT perspective, building on the paradigms and models described in Section \ref{sec:paradigms}. Specifically, it examines the technologies that enable the implementation of these abstract models, focusing on key protocols, software, tools, libraries, and frameworks for distributed computing, communication, and deployment.

\subsection{Computational Frameworks}

\mypar{Distributed Processing}
The execution of batch and stream processing tasks across distributed edge-cloud environments is enabled by specialized computational frameworks. These frameworks typically leverage DAGs to define and manage task dependencies, facilitating optimized scheduling, efficient resource allocation, and robust fault tolerance mechanisms. Popular tools include Apache Spark~\cite{Spark} and Apache Flink~\cite{belcastro2022programming}, both widely adopted for their strong capabilities in handling complex analytics tasks at scale~\cite{DIASDEASSUNCAO20181,rosendo2022distributed}. Apache Spark is particularly well-suited for batch and micro-batch processing, while Apache Flink is primarily used for real-time stream processing. Additionally, Apache Storm~\cite{belcastro2022programming} is another notable framework for real-time streaming analytics, supporting low-latency data processing.

\mypar{Distributed Learning} 
Distributed learning frameworks, including TensorFlow, PyTorch Distributed, and Horovod, have emerged to support large-scale deep learning across multiple machines and GPUs. TensorFlow enables training on heterogeneous systems, from mobile devices to large distributed setups, supporting a wide range of algorithms and applications~\cite{abadi2016tensorflow} while facilitating both data parallelism and model parallelism. It integrates seamlessly with Kubernetes and other cluster management systems (see Section~\ref{sec:deploy_frameworks}), enabling large-scale deployments. Horovod~\cite{sergeev2018horovod}, built on TensorFlow, simplifies distributed training by using efficient inter-GPU communication through ring reduction, reducing communication overheads. Similarly, PyTorch Distributed~\cite{paszke2019pytorch} supports parallel training across different devices and machines, offering various communication backends. These tools abstract much of the complexity of distributed training, allowing developers to focus on model development and experimentation.
Ray~\cite{moritz2018ray} is another framework that simplifies the development of distributed applications, including distributed learning workloads. It provides a unified API for tasks, actors, and distributed objects, enabling efficient parallel execution and resource management. Apache Spark, with its MLlib library, provides scalable machine learning capabilities designed to process large datasets across clusters. MLlib leverages Spark’s in-memory computation engine to optimize performance~\cite{meng2016mllib, 10621567}.

\mypar{Communication-Efficient and Privacy-Preserving Learning}
Several platforms have emerged to support federated learning, each with unique features tailored to different use cases. One of the leading frameworks is TensorFlow Federated (TFF) ~\cite{mcmahan2017communication}, which provides an open-source environment for developers to experiment with various aggregation methods and privacy-preserving techniques. Similarly, FATE (Federated AI Technology Enabler)~\cite{liu2021fate} focuses on federated learning with a strong emphasis on security and privacy, offering a robust platform for building secure FL applications, which makes it particularly suitable for industries that require stringent data protection measures. Another notable framework is Flower~\cite{beutel2020flower}, which is designed to support federated learning across heterogeneous devices. Flower’s flexibility allows for seamless integration across various platforms, making it ideal for environments where device diversity is a key challenge. Recent studies have also demonstrated Flower’s effectiveness in training large language models (LLMs) across diverse computing environments, addressing issues such as device variability, communication efficiency, and scalability~\cite{sani2024future}. IBM Federated Learning (IBM FL)~\cite{ludwig2020ibm} is another prominent platform that provides enterprise-grade federated learning solutions with an emphasis on security, privacy, and compliance. This platform supports various machine learning frameworks and comes equipped with tools for managing data governance. In addition to these platforms, PySyft~\cite{ziller2021pysyft} is a popular library for privacy-preserving machine learning that facilitates federated learning by enabling computations on decentralized data without requiring direct data sharing.
While these frameworks primarily target federated learning, many of them can also be adapted for split learning scenarios. For example, TensorFlow, PyTorch, and PySyft offer the flexibility to define and train model segments on different devices.

\mypar{On-device Machine Learning}
On-device machine learning relies on frameworks like TensorFlow Lite~\cite{abadi2016tensorflow}, which is designed for deploying models on mobile and embedded devices. It optimizes models for size and performance, enabling efficient inference on resource-constrained devices. For TinyML, TensorFlow Lite Micro~\cite{david2021tensorflow} is specifically designed for microcontrollers with extremely limited resources. It supports a subset of TensorFlow operations and is optimized for minimal memory footprint. Edge Impulse~\cite{hymel2022edge} is a platform that simplifies the development and deployment of TinyML applications, offering tools for data collection, model training, and optimization from microcontrollers to gateways. ONNX (Open Neural Network Exchange)~\cite{onnx} is an open format for representing machine learning models that can be used to exchange models between different frameworks and devices. It facilitates the deployment of models on a variety of edge devices. Apache TVM~\cite{chen2018tvm} is a compiler framework for machine learning that optimizes models for different hardware platforms, including edge devices, thus improving the performance of on-device learning models.

\subsection{Communication Protocols}

\mypar{Client-Server} 
A foundational protocol for client-sever communication is HTTP (Hypertext Transfer Protocol), which supports RESTful architectures widely used in cloud environments. REST (Representational State Transfer) enables scalable, stateless interactions between distributed services by leveraging standard HTTP methods, making it a key choice for cloud-native and microservices architectures. With the emergence of HTTP/3, web communication has undergone a significant transformation. Built on QUIC, HTTP/3 offers low latency, seamless network switching, and built-in encryption. A key advantage of QUIC in the edge-cloud continuum is its ability to maintain active connections even as network parameters change. Unlike TCP, which ties connections to a specific IP-port tuple, QUIC uses connection identifiers, allowing sessions to persist across network migrations, wireless handovers, and edge node transitions~\cite{PULIAFITO2022101580}. This makes HTTP/3 particularly effective for IoT, real-time analytics, and mobile applications, ensuring fast, secure, and uninterrupted communication in dynamic cloud-edge environments~\cite{langley2017http3,perna2022first}. Complementing HTTP, CoAP (Constrained Application Protocol)~\cite{coap} is specifically optimized for constrained IoT environments, where lightweight communication is critical. Operating over UDP, CoAP enables request-response interactions that mirror HTTP but with lower overhead, making it ideal for resource-limited devices. Recent advancements include dynamic congestion control mechanisms, enhanced retransmission timeout calculations, and multicast communication capabilities. In particular, extensions such as CoCoA+ (Congestion Control/Advanced)~\cite{betzler2015cocoa+} and secure CoAP variants (e.g., CoAP-DTLS~\cite{kumar2020enhanced}) improve both performance and security, ensuring adaptability while maintaining reliable communication with higher layers. For real-time, persistent client-server communication, WebSocket~\cite{fette2011websocket} provides a full-duplex, event-driven protocol that reduces the overhead of repeated HTTP requests. Its ability to maintain a persistent connection over a single TCP handshake makes it particularly well-suited for applications requiring low-latency updates. Despite its efficiency, WebSocket relies on TCP, which may not be optimal in highly constrained environments, necessitating hybrid approaches with CoAP or MQTT for edge use cases~\cite{websocket-iot-perf}.

\mypar{Publish-Subscribe} 
MQTT (Message Queuing Telemetry Transport) implements the publish-subscribe model with its lightweight, broker-based design~\cite{stanford1999mqtt}. Widely adopted in IoT systems, it ensures efficient data transmission between devices with minimal resource consumption~\cite{light2017mqtt}. Its extensions, such as MQTT-SN (Sensor Networks) and lightweight brokers like Mosquitto, cater specifically to constrained devices~\cite{dizdarevic2019iot}. Additionally, adaptations like MQTT-ST (Spanning Tree) enhance routing and failure recovery, ensuring scalability for large IoT networks~\cite{longo2019mqttst}. Integration with modern transport protocols like QUIC further reduces connection overhead, improving MQTT’s performance in high-latency environments~\cite{kumar2019mqttquic}.For more complex middleware messaging needs, AMQP (Advanced Message Queuing Protocol) introduces advanced features such as message persistence, transactionality, and routing~\cite{vinoski2006amqp}, particularly useful in distributed enterprise applications and IoT ecosystems where~\cite{williams2012rabbitmq}.  Although it demands higher resource consumption than lightweight alternatives, AMQP is a preferred choice in fog-to-cloud deployments~\cite{yakupov2022overview}.
A further model based on the publish-subscribe paradigm is XMPP (Extensible Messaging and Presence Protocol), which provides a structured and extensible XML-based communication standard~\cite{saint2011extensible}. XMPP has evolved to support publish-subscribe interactions in IoT and edge-cloud environments, optimizing message formats to reduce energy consumption in resource-constrained devices, despite its reliance on verbose XML. DDS (Data Distribution Service)~\cite{pardo2016dds} also follows the publish-subscribe paradigm but is specifically designed for real-time, scalable, and high-performance data exchange. Unlike MQTT or AMQP, DDS employs a decentralized peer-to-peer model where nodes communicate directly via UDP/IP unicast and TCP/IP multicast, reducing dependency on central brokers. This makes it particularly effective for large-scale IoT applications where low-latency and high-throughput communication are essential. DDS also incorporates security mechanisms such as TLS and DTLS, but its decentralized nature also introduces challenges, such as increased susceptibility to DoS and DDoS attacks~\cite{fi12030055}. Recent developments focus on integrating AMQP with other publish-subscribe protocols like MQTT and DDS, enhancing interoperability across heterogeneous systems~\cite{dizdarevic2019iot}.

\mypar{Actors} 
While no specific protocol fully embodies the actor model in its pure form, some messaging protocols can be adapted for actor-based architectures. Systems using AMQP or MQTT can implement actor-like behavior by ensuring each entity processes messages independently and asynchronously, without shared state. Moreover, actor model-based frameworks have been proposed for the edge-cloud continuum. Among these, Akka Edge~\cite{akka} is a toolkit for building concurrent, distributed, and resilient message-driven applications, which allows for developing scalable and fault-tolerant systems in the edge-cloud continuum~\cite{srirama2021akka}. Similarly, CANTO is a distributed fog framework for training neural networks in IoT applications, addressing latency issues by processing data closer to edge device~\cite{srirama2023canto}.  

\subsection{Deployment Frameworks}
\label{sec:deploy_frameworks}

\mypar{Virtualization}
In the context of the edge-cloud continuum, where computational resources span from centralized data centers to distributed edge nodes, the choice of hypervisor for server virtualization plays a critical role. Proprietary hypervisors such as VMware ESXi and Microsoft Hyper-V have traditionally dominated enterprise data centers due to their mature management ecosystems and integration with enterprise software stacks. However, these solutions are often considered too rigid and resource-intensive for edge environments, where hardware is limited and operational simplicity is key. 
In contrast, KVM, an open source hypervisor directly integrated into the Linux kernel, has emerged as a more flexible and lightweight alternative. Its minimal overhead, native compatibility with cloud-native orchestration tools like Kubernetes and OpenStack, and its ability to be deeply customized make it particularly suitable for edge deployments~\cite{raho2015kvm,Soriga2013ACO}.
Xen is another open-source hypervisor that once played a leading role in early cloud platforms and served as the original hypervisor used by Amazon AWS before transitioning to its custom KVM-based Nitro hypervisor~\cite{schlaeger2018aws,Deshane2008QuantitativeCO}. While Xen still finds application in certain niche environments, such as telecommunications and some real-time systems, its broader adoption has declined in favor of KVM. As a result, within the edge-cloud continuum, KVM has become the preferred hypervisor for lightweight, scalable, and cost-effective edge infrastructure, while ESXi and Hyper-V maintain their strong presence in traditional enterprise data centers.

\mypar{Containerization} The complexity of managing distributed resources and applications necessitates robust orchestration solutions. Orchestrators automate the deployment, scaling, and management of containerized applications, ensuring efficient resource utilization, fault tolerance, and high availability across diverse environments. Tools like Kubernetes\footnote{https://kubernetes.io} and Apache Mesos\footnote{https://mesos.apache.org/} can manage VMs, microVMs, and containers, making it easy to deploy, scale, and manage applications in edge-cloud infrastructures. Some of these tools ensure resilience and reliability by automating the detection and recovery from various types of failures (i.e., \virgolette{self-healing}), which reduces the need for manual intervention. 
A brief comparison of the most popular orchestration tools is shown in Table~\ref{tab:comparison-container-orchestrators}.

\begin{table}[!htb]
\fontsize{7pt}{7pt}\normalfont{
\centering
\begin{tabular}{@{}llllllll@{}}
\toprule
\textbf{Feature}             & \textbf{Kubernetes}     & \textbf{Docker Swarm} & \textbf{Apache Mesos} & \textbf{Nomad~\footnote{https://www.nomadproject.io/}} & \textbf{OpenShift~\footnote{https://www.redhat.com/en/technologies/cloud-computing/openshift}} & \textbf{Rancher~\footnote{https://www.rancher.com/}} \\ \midrule
\textbf{Scalability}         & High                    & Moderate              & High                  & High           & High               & High                \\[7pt]
\textbf{Ease of Use}         & Medium                & High                  & Medium              & Medium       & Medium           & High                      \\[7pt]
\textbf{Resource Management} & Advanced                & Basic                 & Advanced              & Advanced       & Advanced           & Medium                 \\[7pt]
\textbf{Fault Tolerance}     & Yes                     & Yes                   & Yes                   & Yes            & Yes                & Yes                        \\[7pt]
\textbf{Extensibility}       & High                    & Medium              & High                  & High           & High               & Medium             \\[7pt]
\textbf{Scalability}     & High         & Small       & High       & High & High   & Medium  \\[7pt]
\textbf{Self-Healing}        & Yes                     & Basic                 & Yes                   & Yes            & Yes                & Yes                     \\[7pt]
\textbf{Support for VMs}     & \begin{tabular}[c]{@{}l@{}}Yes (via KubeVirt) \end{tabular}    & No                    & Yes                   & No             & \begin{tabular}[c]{@{}l@{}}Yes (via KubeVirt) \end{tabular} & No                       \\[7pt]
\textbf{Support for MicroVMs} & \begin{tabular}[c]{@{}l@{}}Yes (via Firecracker) \end{tabular}  & No                    & No                    & No             & \begin{tabular}[c]{@{}l@{}}Yes (via Firecracker) \end{tabular}  & No                  \\[7pt]
\textbf{Container Formats}   & \begin{tabular}[c]{@{}l@{}}Docker, CRI-O,\\ containerd \end{tabular}              & Docker, Mesos         & Docker         & 

  \begin{tabular}[c]{@{}l@{}}Docker, CRI-O,\\ containerd \end{tabular}
& Docker & Docker        \\ \bottomrule
\end{tabular}
\caption{Comparison of container orchestration tools.}
\label{tab:comparison-container-orchestrators}
}
\end{table}

These tools are often delivered through cloud-based services to efficiently manage containerized applications, following a Container-as-a-Service (CaaS) paradigm. With CaaS, organizations can deploy, orchestrate, and scale containers without managing the underlying infrastructure, allowing developers to focus on application development rather than operational complexities. CaaS platforms from leading cloud providers include Amazon Elastic Container Service (ECS) and Elastic Kubernetes Service (EKS), Google Kubernetes Engine (GKE), Microsoft Azure Kubernetes Service (AKS), Alibaba Cloud Container Service for Kubernetes, and IBM Cloud Kubernetes Service. In terms of orchestration, these services rely almost exclusively on Kubernetes, which is the lead solution for managing complex and large scale deployments. 

\mypar{Serverless}
All major cloud vendors provide developers with their own FaaS services (e.g., AWS Lambda, Azure Functions, Google Cloud Functions). However, several other open-source frameworks have been developed to cope with the different requirements, such as geographical distribution, decentralized scheduling, function offloading, live function migration, and function composition. Table~\ref{tab:comparison-faas} shows an overview of the most popular FaaS frameworks and their features, including Kubedge~\cite{Xiong2018ExtendCT}, Colony~\cite{lordan2021colony}, and Serverledge~\cite{RUSSORUSSO2024101915}. Several other FaaS frameworks have been proposed, but they do not allow deploying and managing functions across multiple geographic locations. This capability is crucial in the context of the edge-cloud continuum to enable placement and management of computing resources across both cloud data centers and edge locations.
Examples of FaaS framework that do not support geographical distribution are OpenWhisk\footnote{https://openwhisk.apache.org/}, OpenFaaS\footnote{https://www.openfaas.com/}, and tinyFaaS~\cite{pfandzelter2020tinyfaas}. 

\begin{table}[!htb]
\fontsize{7pt}{7pt}\normalfont{
\centering
\begin{tabular}{@{}lcccccccl@{}}
\toprule
\textbf{Framework} & \textbf{Distrib.} & \textbf{\begin{tabular}[c]{@{}l@{}} Scheduling\end{tabular}} & \textbf{Offloading} & \textbf{\begin{tabular}[c]{@{}l@{}}Live\\ Migration\end{tabular}} & \textbf{\begin{tabular}[c]{@{}l@{}}Function\\ Composit.\end{tabular}} & \textbf{Runtime} & \textbf{\begin{tabular}[c]{@{}l@{}}Latency\end{tabular}} & \textbf{\begin{tabular}[c]{@{}l@{}}Supported \\ Languages\end{tabular}} \\ \midrule
\textbf{Colony} & Geo & D & yes & - & yes & COMPSs & Low & \begin{tabular}[c]{@{}l@{}}Java, Python, C++\end{tabular} \\[7pt]
\textbf{funcX} & Geo & C & - & - & yes & \begin{tabular}[c]{@{}l@{}}Node or\\  Containers\end{tabular} & Medium & Python \\[7pt]
\textbf{Serverledge} & Geo & D & yes & yes & - & Containers & Low & \begin{tabular}[c]{@{}l@{}}Python, Node.js, any\end{tabular} \\[7pt]

\textbf{OpenWhisk} & Local & C & - & - & yes & Containers & Low & \begin{tabular}[c]{@{}l@{}}Go, Java, NodeJS, .NET,\\ PHP, Python, Ruby,\\Rust, Scala, Swift\end{tabular} \\ [14pt]

\textbf{OpenFaaS} & Local & C & - & - & yes & Containers & Low & \begin{tabular}[c]{@{}l@{}}    Go, Node.js, Python, C\# \end{tabular} \\[7pt]
\textbf{tinyFaaS} & Local & C & - & - & - & Containers & Low & \begin{tabular}[c]{@{}l@{}}Go, Node.js, Python, binary\end{tabular} \\[7pt]
\textbf{\begin{tabular}[c]{@{}l@{}}AWS \\Lambda\end{tabular}} & CDN & C & - & no & yes & MicroVMs & High & \begin{tabular}[c]{@{}l@{}}Python, Node.js, C\#,\\ Java, Ruby, Go\end{tabular} \\
[7pt]
\textbf{\begin{tabular}[c]{@{}l@{}}Google \\Cloud\\Functions\end{tabular}} & CDN & C & - & no & yes & Containers & Medium & \begin{tabular}[c]{@{}l@{}}Python, Node.js, Go, Java\end{tabular} \\[7pt]
\textbf{\begin{tabular}[c]{@{}l@{}}Azure\\  Functions\end{tabular}}& CDN & C & - & no & yes & Containers & Medium & \begin{tabular}[c]{@{}l@{}}C\#, Java, Python, Javascript, \\Typescript, Powershell\end{tabular} \\ [7pt]
\textbf{\begin{tabular}[c]{@{}l@{}}KubeEdge\end{tabular}} & Local & C & yes & no & yes & Containers & Low & \begin{tabular}[c]{@{}l@{}}Python, Go, Java, Node.js\end{tabular} \\
\bottomrule
\end{tabular}
\caption{Comparison of FaaS frameworks and cloud-based FaaS services (D=decentralized, C=centralized).}
\label{tab:comparison-faas}
}
\end{table}

\section{Deployment Platforms}
\label{sec:platforms}

This section explores the major cloud platforms that enable computation and data management across the edge–cloud continuum, which can be broadly classified into public and private platforms. Public cloud solutions, such as Amazon Web Services (AWS), Microsoft Azure, and Google Cloud Platform (GCP), deliver resources over the internet to multiple customers, providing scalability and cost-efficiency. In contrast, private clouds, typically based on open-source solutions such as OpenStack or OpenNebula, are deployed within an organization’s infrastructure, providing greater control, customization, and data sovereignty. 
Hybrid and multi-cloud approaches can improve service delivery by balancing on-premises and cloud resources. They enable low-latency processing closer to users, while leveraging the scalability of cloud services for advanced analytics. Such strategies improve resilience by distributing workloads, minimizing vendor lock-in, and increasing fault tolerance. Some hybrid edge-cloud frameworks support this by optimizing service across public and private clouds, either by sharing resources or maintaining physical isolation between them~\cite{gu2023hyedge, alamouti2022hybrid}.

Beyond general-purpose cloud services, cloud platforms also provide \textit{edge-specific enabling services} that extend cloud-like capabilities to different layers of the continuum, allowing organizations to run applications seamlessly at the edge while integrating with core cloud offerings like AI/ML, analytics, and orchestration services.
These services act as a \textit{bridge} for running standard cloud offerings in environments outside the central data center. 
For example, \textit{AWS Outposts} brings AWS compute (EC2) and other AWS services onto a dedicated server that resides in an on-premise data center. Similarly, \textit{Azure Stack Edge} provides managed local devices bringing Azure services (e.g., compute, storage, AI) to the edge. Other solutions, such as the \textit{AWS Snow Family} or \textit{Huawei Intelligent EdgeFabric (IEF)}, specialize in bringing compute and storage closer to the data source, particularly in challenging environments (e.g., remote locations with limited connectivity). Table~\ref{tab:edge_cloud_services} provides an overview of the enabling services provided by major cloud platforms for each level of the computing continuum. Further details on these services are discussed in the next sections.

\begin{table}[htb!]
\centering
\fontsize{7pt}{7pt}\normalfont{
\begin{tabular}{@{}lcccc@{}}
\toprule
\textbf{Platform} 
 & \textbf{Near Edge} 
 & \textbf{Far Edge} 
 & \textbf{On-Premise} 
 & \textbf{On-Device} \\ 
\midrule

\textbf{AWS} 
 & \begin{tabular}[c]{@{}c@{}}AWS Local Zones\\AWS Snowball Edge\end{tabular} 
 & \begin{tabular}[c]{@{}c@{}}AWS Wavelength\\AWS Snowcone\end{tabular} 
 & \begin{tabular}[c]{@{}c@{}}AWS Outposts\end{tabular} 
 & \begin{tabular}[c]{@{}c@{}}AWS IoT Greengrass\\AWS IoT Core\end{tabular}
\vspace{0.2cm}
\\

\textbf{Azure} 
 & \begin{tabular}[c]{@{}c@{}}Azure Private MEC\end{tabular}
 & \begin{tabular}[c]{@{}c@{}}Azure Edge Zones \\(for telco operators)\end{tabular}
 & \begin{tabular}[c]{@{}c@{}}Azure Stack Edge\\Azure Stack HCI\end{tabular}
 & \begin{tabular}[c]{@{}c@{}}Azure IoT Edge\end{tabular}
\vspace{0.2cm}
\\

\textbf{Google Cloud} 
 & \begin{tabular}[c]{@{}c@{}}Google Distributed Cloud Edge\end{tabular}
 & \begin{tabular}[c]{@{}c@{}}Google Distributed Cloud Edge\end{tabular}
 & \begin{tabular}[c]{@{}c@{}}Google Distributed Cloud Hosted\\Anthos on-prem\end{tabular}
 & \begin{tabular}[c]{@{}c@{}}Google Coral Edge TPU\end{tabular}
\vspace{0.2cm}
\\

\textbf{IBM Cloud} 
 & \begin{tabular}[c]{@{}c@{}}IBM Edge Application\\Manager\end{tabular}
 & \begin{tabular}[c]{@{}c@{}}IBM Edge Application\\Manager\end{tabular}
 & \begin{tabular}[c]{@{}c@{}}IBM Cloud Satellite\end{tabular}
 & \begin{tabular}[c]{@{}c@{}}--\end{tabular}
\vspace{0.2cm}
\\

\textbf{Alibaba Cloud} 
 & \begin{tabular}[c]{@{}c@{}}Link IoT Edge\\Edge Nodes (regional)\end{tabular}
 & \begin{tabular}[c]{@{}c@{}}Link IoT Edge\end{tabular}
 & \begin{tabular}[c]{@{}c@{}}Apsara Stack\end{tabular}
 & \begin{tabular}[c]{@{}c@{}}Link IoT Platform\end{tabular}
\vspace{0.2cm}
\\

\textbf{Huawei Cloud} 
 & \begin{tabular}[c]{@{}c@{}}Intelligent EdgeFabric (IEF)\\Huawei MEC\end{tabular}
 & \begin{tabular}[c]{@{}c@{}}Intelligent EdgeFabric (IEF)\\EdgeGallery (open source)\end{tabular}
 & \begin{tabular}[c]{@{}c@{}}Huawei Cloud Stack\\FusionCube\end{tabular}
 & \begin{tabular}[c]{@{}c@{}}LiteOS\end{tabular}
\vspace{0.2cm}
\\

\textbf{Tencent Cloud} 
 & \begin{tabular}[c]{@{}c@{}}EdgeOne (CDN/security)\\(MEC pilots)\end{tabular}
 & \begin{tabular}[c]{@{}c@{}}EdgeOne\\(with telco partners)\end{tabular}
 & \begin{tabular}[c]{@{}c@{}}Tencent Cloud TCE\end{tabular}
 & \begin{tabular}[c]{@{}c@{}}Tencent IoT Explorer\end{tabular}
\vspace{0.2cm}
\\

\textbf{OpenStack} 
 & \begin{tabular}[c]{@{}c@{}}StarlingX\end{tabular}
 & \begin{tabular}[c]{@{}c@{}}StarlingX\end{tabular}
 & \begin{tabular}[c]{@{}c@{}}OpenStack\end{tabular}
 & \begin{tabular}[c]{@{}c@{}}StarlingX\end{tabular}
\vspace{0.2cm}
\\

\textbf{OpenNebula} 
 & \begin{tabular}[c]{@{}c@{}}OpenNebula Edge\end{tabular}
 & \begin{tabular}[c]{@{}c@{}}(Community extensions)\end{tabular}
 & \begin{tabular}[c]{@{}c@{}}OpenNebula\end{tabular}
 & \begin{tabular}[c]{@{}c@{}}--\end{tabular}
\\
\bottomrule
\end{tabular}
}
\caption{
A comparison of key services offered by major cloud platforms across the different layers of the compute continuum.}
\label{tab:edge_cloud_services}
\end{table}

\subsection{Public Platforms}

\mypar{Amazon Web Services (AWS)}
At the near edge, \textit{AWS Local Zones} brings AWS services closer to major metropolitan areas for low-latency applications, while \textit{AWS Snowball Edge} provides data processing and migration for remote locations within rugged appliances. At the far edge, \textit{AWS Wavelength} integrates with 5G networks by embedding AWS compute and storage services within telecommunications infrastructure. \textit{AWS Snowcone} provides portable edge devices for remote or mobile deployments. On-premise solutions leverage \textit{AWS Outposts} to enable organizations running AWS infrastructure and services in their own data centers. Finally, for on-device capabilities, \textit{AWS IoT Greengrass} and \textit{AWS IoT Core} enable developers to deploy and manage IoT applications, execute local computing tasks and edge AI models.

\mypar{Microsoft Azure}
At the near edge, \textit{Azure Private MEC} enables enterprises to deploy low-latency, high-performance applications by integrating Azure cloud services with private 5G or LTE networks. At the far edge, \textit{Azure Edge Zones} bring Azure services close to metro areas or telecom networks, leveraging 5G connectivity and carrier partnerships.
For on-premise needs, \textit{Azure Stack Edge} and \textit{Azure Stack HCI} deliver Azure services in local data centers. At the on-device layer, \textit{Azure IoT Edge} facilitates local data processing, AI, and device management.

\mypar{Google Cloud}
At the near edge and far edge, \textit{Google Distributed Cloud Edge} integrates telecom 5G networks to deliver low-latency compute resources in enterprise facilities.
For on-premise deployments, \textit{Google Distributed Cloud Hosted} and \textit{Anthos on-prem} provide an isolated or hybrid environment that uses Google Cloud-based Kubernetes infrastructure for managing workloads across cloud and on-premise infrastructure. For on-device solutions, \textit{Google Coral Edge TPU} offers specialized hardware accelerators that enable efficient AI inference on low-power devices.

\mypar{IBM Cloud}
IBM provides enterprise-focused edge solutions. At the near edge, \textit{IBM Edge Application Manager} autonomously manages distributed edge workloads across multiple edge sites. The same platform applies to far edge locations, helping orchestrate containerized services on nodes with limited connectivity.
For on-premise environments, \textit{IBM Cloud Satellite} provides a managed distributed cloud that extends IBM Cloud services into customer data centers or private infrastructures. IBM solutions do not currently include a dedicated on-device operating system or SDK, leaving on-device responsibilities to third-party or community tooling.

\mypar{Alibaba Cloud}
Alibaba provides a broad range of edge computing solutions for industrial IoT and hybrid cloud scenarios, particularly suited to users in the Asia-Pacific region. Near edge solutions include \textit{Link IoT Edge}, which supports local data processing, and regional \textit{Edge Nodes} for accelerating content and compute. At the far edge, the same \textit{Link IoT Edge} service can be deployed on smaller remote gateways or devices to handle low-latency tasks closer to data sources. On-premise customers can use \textit{Apsara Stack} as a hybrid solution to run Alibaba Cloud services within their own data centers. Finally, the on-device layer is supported by the \textit{Link IoT Platform} and \textit{Device SDK}, enabling device-to-cloud connectivity, data ingestion, and remote device management.

\mypar{Huawei Cloud}
Huawei Cloud delivers a range of edge-oriented tools and frameworks, often coupled with telco partners. Near edge capabilities revolve around \textit{Intelligent EdgeFabric (IEF)} and \textit{Huawei MEC}, which bring compute and storage resources to base stations or local points-of-presence. At the far edge, the same \textit{IEF} platform extends into remote environments, and the open source \textit{EdgeGallery} service supports developing and deploying edge applications on 5G networks. On-premise scenarios are addressed by \textit{Huawei Cloud Stack} and \textit{FusionCube}, providing private cloud infrastructure that integrates with Huawei Cloud services. For on-device, Huawei offers \textit{LiteOS}, a lightweight real-time operating system, and an \textit{IoT Device SDK} to build and connect embedded devices securely.

\mypar{Tencent Cloud}
Tencent Cloud delivers edge services with a strong emphasis on content delivery and early-stage MEC deployments. At the near edge, \textit{EdgeOne} serves as a global CDN to minimize latency for content delivery. Extending to the far edge, the \textit{EdgeOne} backbone brings compute resources closer to users, enabling low-latency applications across geographically distributed regions. For on-premise deployments, \textit{Tencent Cloud TCE} provides private cloud solutions within customer data centers. At the on-device layer, \textit{Tencent IoT Explorer} and its SDK support IoT connectivity and application development from edge devices to the Tencent cloud.

\vspace{0.5em}
These cloud platforms maintain distinct global infrastructures, with AWS, Microsoft Azure, Google Cloud, and Alibaba representing the primary providers in terms of scale and geographic coverage. Figure~\ref{fig:cloud-infrastructures} illustrates these differences by highlighting their respective cloud regions and near-edge zones.

\definecolor{cloudregion}{RGB}{1, 102, 171}         
\definecolor{localzone}{RGB}{195, 61, 105}          
\begin{figure}[htb!]
    \centering
    \begin{subfigure}{0.49\textwidth}
        \centering
        \includegraphics[width=\textwidth]{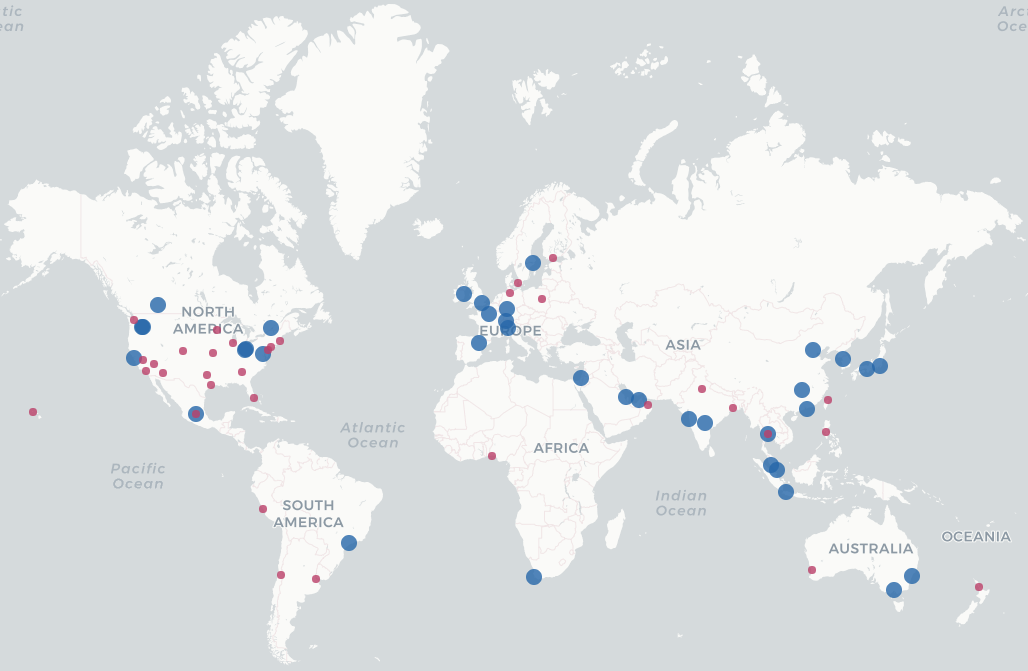}
        \caption{Amazon Web Services:
                \textcolor{cloudregion}{\Large $\bullet$} (36), \textcolor{localzone}{\Large $\bullet$} (34) }
        \label{fig:sub1}
    \end{subfigure}
    \hfill
    \begin{subfigure}{0.49\textwidth}
        \centering
        \includegraphics[width=\textwidth]{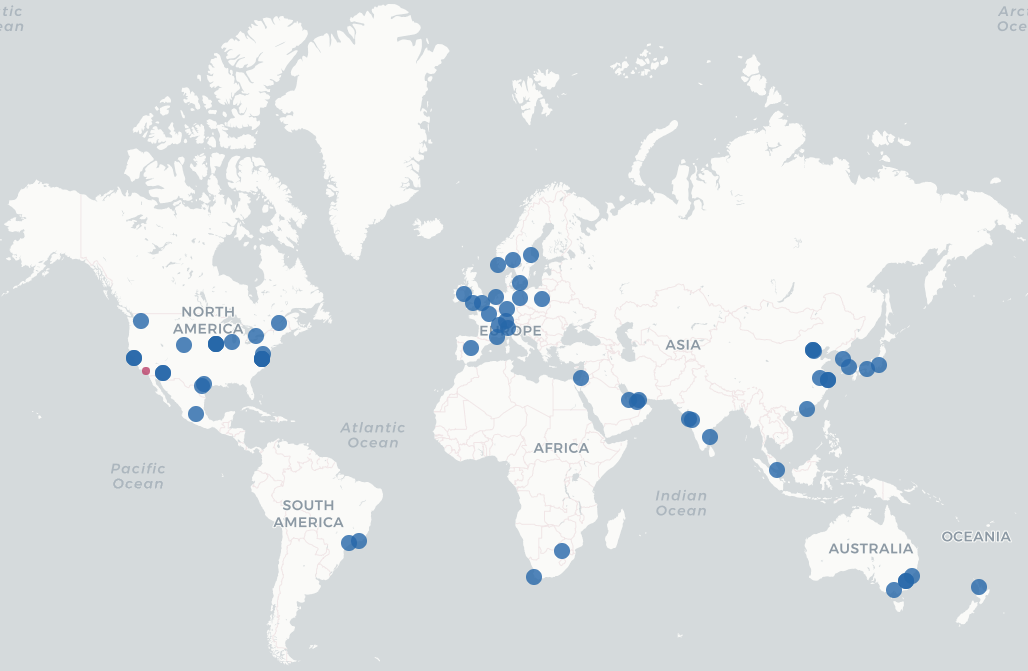}
        \caption{Microsoft Azure: 
        \textcolor{cloudregion}{\Large $\bullet$} (65), \textcolor{localzone}{\Large $\bullet$} (1) }
        \label{fig:sub2}
    \end{subfigure}
    \vskip\baselineskip
    \begin{subfigure}{0.49\textwidth}
        \centering
        \includegraphics[width=\textwidth]{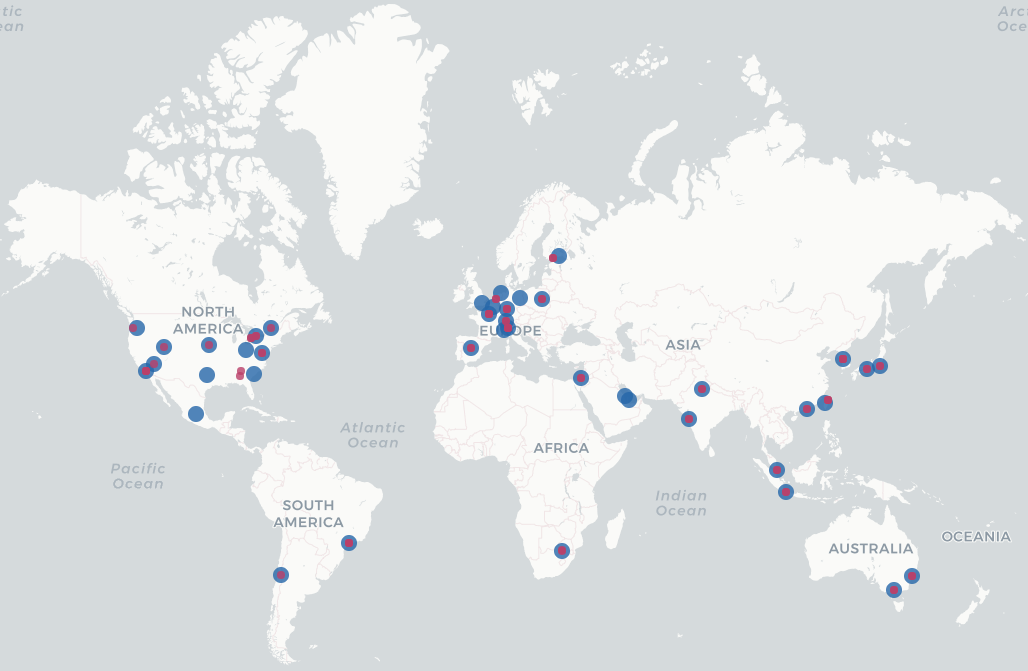}
        \caption{Google Cloud:
                \textcolor{cloudregion}{\Large $\bullet$} (41), \textcolor{localzone}{\Large $\bullet$} (71) }
        \label{fig:sub3}
    \end{subfigure}
    \hfill
    \begin{subfigure}{0.49\textwidth}
        \centering
        \includegraphics[width=\textwidth]{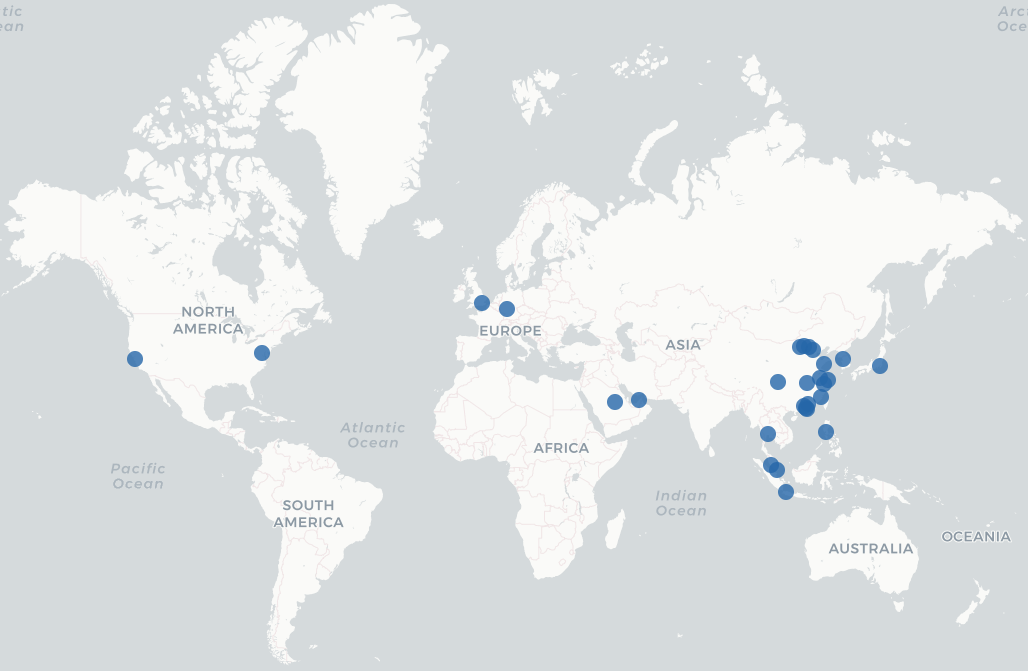}
        \caption{Alibaba:         
        \textcolor{cloudregion}{\Large $\bullet$} (28), 
        \textcolor{localzone}{\Large $\bullet$} (0) }
        \label{fig:sub4}
    \end{subfigure}
    \caption{Global infrastructure of major cloud providers, including the number of active cloud regions and local zones for each provider. Source: \url{https://www.cloudinfrastructuremap.com/} [updated to February, 2025]. 
    \textbf{Legend:} 
    \textcolor{cloudregion}{\Large $\bullet$} Cloud Regions, 
    \textcolor{localzone}{\Large $\bullet$} Near-Edge Zones.}
    \label{fig:cloud-infrastructures}
\end{figure}

Specifically, Microsoft Azure maintains the most extensive network of cloud regions, with 65 globally distributed sites. In comparison, Amazon Web Services (AWS), while operating a smaller number of cloud regions (36), demonstrates a more balanced deployment strategy through substantial investment in near-edge zones (34), particularly in urban centers such as Miami, Berlin, and Seoul, thereby supporting low-latency services. Google Cloud adopts a similarly edge-focused approach, with 41 cloud regions and the largest number of near-edge zones (71), strategically located in key metropolitan areas including Los Angeles, London, Tokyo, and São Paulo, reflecting a prioritization of high-performance, proximity-based computing. In contrast, Alibaba Cloud concentrates its infrastructure within a more limited geographic scope, primarily across the Asia-Pacific region, with 28 cloud regions and no near-edge zones.

\subsection{Private Platforms}

\mypar{OpenStack}
OpenStack provides a popular open-source framework for diverse edge computing scenarios. \textit{StarlingX}, an open-source edge computing platform built on OpenStack and Kubernetes, supports both near edge and far edge deployments. OpenStack also supports on-premise deployments, allowing organizations to manage virtual machines, storage, and networks in private data centers while providing a flexible approach to hybrid cloud infrastructure. Additionally, \textit{StarlingX} can be adapted for on-device deployments, though it is primarily used for gateways and micro data centers, rather than resource-constrained devices. 

\mypar{OpenNebula}
OpenNebula is a lightweight, open-source alternative to OpenStack. 
At the near edge, \textit{OpenNebula Edge} allows orchestration and resource allocation across distributed edge nodes. Far edge support relies on community extensions, often tailored for low-resource hardware in remote locations. For example, OpenNebula ONEedge5G is a recent industrial research initiative focused on enabling efficient, automated deployment of distributed edge environments over 5G infrastructures by integrating AI techniques and easy resource management. For on-premise deployments, OpenNebula provides a traditional private cloud management toolkit, enabling enterprises to run cloud environments within their data centers. Unlike OpenStack, OpenNebula does not offer an official on-device solution, leaving this area to external or community-driven projects. 

\section{Applications Across the Continuum}
\label{sec:application-domains}

It is important to consider key application domains where edge-cloud integration delivers significant advantages. Particularly, here we describe use cases in healthcare, industrial IoT, smart cities, and real-time services, discussing how distributed computing enhances performance and efficiency. Additionally, we provide an overview of benchmarking tools used to evaluate these systems, offering insights into their capabilities and trade-offs.

\subsection{Application Domains}

\mypar{Healthcare}
Healthcare has benefited greatly from the integration of edge computing to enable local data processing and immediate insights~\cite{Pace2019AnEA, Aujla2019SAFESF}. Applications span remote patient monitoring, real-time health data analytics, and smart medical devices~\cite{sanchez2020continuous, tuli2020healthfog, rahmani2018exploiting}. By offloading computation closer to the data source (e.g., at hospital gateways or local clinical servers), the edge-cloud continuum reduces latency, allowing faster diagnostic results and timely interventions in critical situations. For instance, frameworks like \textit{HealthEdge}\footnote{https://healthedge.com/} use edge servers to predict complications (e.g., diabetes) on a per-patient basis, improving care efficiency by delivering real-time notifications and treatment recommendations. Meanwhile, cloud resources can centrally aggregate large-scale medical data for advanced analytics, longitudinal studies, and system-wide optimization, often leveraging federated learning systems to enable privacy-preserving analysis~\cite{bochicchio2024personalized}.

\mypar{Industrial IoT (IIoT)}
In industrial IoT, the edge-cloud continuum optimizes manufacturing processes, predictive maintenance, and real-time equipment monitoring~\cite{Chen2018EdgeCI}. Edge computing on the factory floor reduces latency and ensures immediate response to machinery faults or anomalies, enhancing operational efficiency~\cite{9387301}. Local data analysis enables companies to detect performance bottlenecks and address technical issues in near real time, preventing critical failures~\cite{Zhang2019SeriousCA}. Simultaneously, cloud services aggregate metrics from multiple facilities or lines of production, supporting deeper analytics, fleet-wide pattern recognition, and business intelligence.

\mypar{Smart Cities}
Smart city environments increasingly rely on massive numbers of sensors and Internet-connected devices to manage infrastructure such as traffic systems, energy grids, and public safety networks. With the edge-cloud continuum, data-intensive tasks are distributed across urban gateways and edge servers, allowing real-time applications like adaptive traffic lights~\cite{Mora2019DistributedAF}, environmental monitoring for sustainability~\cite{Khan2019EdgeComputingEnabledSC}, and smart transportation~\cite{edge-cloud-2023-ieee-access}. Local edge processing reduces response times for immediate actions, e.g., redirecting traffic flow or alerting first responders, while the cloud layer focuses on macro-level insights and long-term urban planning. 

\mypar{Real-Time Services}
Real-time services, particularly online gaming and entertainment, require ultra-low latency for a smooth user experience~\cite{Bilal2017EdgeCF}. By processing and caching content closer to users, edge computing mitigates round-trip delays to the cloud, reducing lag and delivering fluid gameplay~\cite{Plumb2018ExploitingGE}. For media streaming, content delivery networks (CDNs) leverage edge caching to deliver high-quality media streams with minimal buffering or disruptions, while the cloud provides centralized orchestration, content management, and analytics at scale~\cite{gama2021video}.

\subsection{Benchmarking Tools}
\label{sec:benchmarking}
Evaluating the performance, reliability, and scalability of edge-cloud continuum systems is challenging due to their geographically distributed nature and heterogeneity~\cite{maheshwari2018scalability}. This complexity necessitates a multifaceted approach to system evaluation. This section analyzes three primary evaluation methods: simulators, emulators, and tests on real architectures, discussing available software frameworks and highlighting their respective advantages and disadvantages.

\mypar{Simulators} Simulators model the behavior and interactions of edge-cloud architectures without deploying actual hardware, providing a controlled environment to efficiently test various configurations. They are cost-effective as they eliminate the need for expensive hardware, reducing experimental costs and offering scalability.  Additionally, they support reproducible research under identical conditions. Simulations facilitate rapid prototyping and testing of new algorithms, protocols, and architectures without the risk of hardware failures. However, the accuracy of simulation results depends on the validity of underlying models, which may not fully capture real-world dynamics, leading to discrepancies when transitioning to deployment. One of the most widely adopted simulation tools is iFogSim~\cite{gupta2017ifogsim, mahmud2022ifogsim2}, an extension of CloudSim~\cite{calheiros2011cloudsim} designed for modeling fog computing infrastructures by considering factors such as network congestion and latency. CloudSimSDN~\cite{cloudsimsdn}, another extension of CloudSim, introduces support for software-defined networking (SDN), allowing more flexible network topology configurations and evaluations of workload distribution strategies. A more recent alternative is YAFS (Yet Another Fog Simulator)~\cite{lera2019yafs}, a Python-based simulator that allows for dynamic topology modeling, analysis of network performance, and adaptive resource allocation strategies. Finally, EdgeCloudSim~\cite{sonmez2018edgecloudsim} offers features such as mobility modeling and network link characterization. Various survey studies offer comprehensive analyses of simulators for edge-cloud environments~\cite{svorobej2019simulating, kumar2021fog, ashouri2019edge}.

\mypar{Emulators} Emulators mimic the behavior of edge-cloud architectures more closely than simulators by running actual software on virtualized or containerized environments that replicate the target hardware. Prominent emulation tools include Mininet and EmuEdge.
Mininet~\cite{kaur2014mininet} is a popular network emulator that creates a virtual network of hosts, switches, and links on a single machine, allowing researchers to prototype large-scale network topologies and test network protocols in a controlled environment. Another widely used emulator is EmuFog~\cite{mayer2017emufog}, which focuses on fog computing scenarios. EmuFog allows researchers to deploy and test applications on a virtual fog infrastructure, providing insights into the performance and scalability of fog-based solutions. Beyond these, the iContinuum toolkit~\cite{Akbari2024iContinuumAE} facilitates intent-based testing and experimentation across the edge-cloud continuum, leveraging SDNs and containerization.
Generally, emulators provide a more accurate representation of real-world performance, as they execute actual software in environments that closely resemble the intended deployment conditions, replicating resource constraints and network conditions. However, emulating complex systems demands substantial computational and memory resources, making it less scalable than simulation.

To effectively evaluate edge-cloud architectures, researchers often use a combination of simulators, emulators, and small-scale test deployments on real systems, leveraging the strengths of each method while mitigating their weaknesses. Simulators are well-suited for initial prototyping and exploration of different configurations. Emulators bridge the gap between abstract models and real-world deployments, providing more accurate performance insights while maintaining some level of flexibility and cost-effectiveness. Finally, tests on real architectures are necessary for final validation and understanding of operational challenges, ensuring that the systems perform as expected in actual deployment scenarios.

\subsection{Application Maintenance}

Maintaining applications in the edge-cloud continuum poses challenges due to the highly distributed and heterogeneous nature of this environment. The wide range of devices requires robust solutions for logging, CI/CD (Continuous Integration/Continuous Deployment), and monitoring. These tools have evolved significantly to address the demands of the edge-cloud continuum, ensuring interoperability and scalability across diverse systems. The following sections explore the state-of-the-art tools and approaches for logging, CI/CD, and monitoring.

\mypar{Logging} Logging in the compute continuum requires aggregating and analyzing decentralized logs across multiple locations without overloading central systems. Tools such as the ELK Stack (Elasticsearch, Logstash, Kibana)\footnote{\url{https://www.elastic.co}} provide centralized log collection, real-time visualization, and customizable dashboards, making them ideal for scalable cloud environments. In contrast, Fluentd\footnote{\url{https://fluentd.org/}} offers a lightweight, pluggable approach suited for resource-constrained edge devices with seamless cloud integration. Managed solutions like \textit{Amazon CloudWatch Logs} and \textit{Azure Monitor Logs} simplify cloud-native logging with built-in scalability, while open-source tools like Graylog\footnote{\url{https://www.graylog.org/}} provide flexible log management for hybrid setups.  
Such logging systems differ in terms of scalability, real-time processing, support for edge devices, and cloud integration, affecting their suitability for the edge-cloud continuum. The ELK Stack is situable for large-scale centralized logging but is resource-intensive, making it less ideal for edge setups. On the contrary, Fluentd is lighter, supports AWS and Azure integration, and works well across heterogeneous environments. \textit{Amazon CloudWatch Logs} and \textit{Azure Monitor Logs} provide real-time processing and scalability but are tightly coupled with their respective cloud platforms, reducing flexibility in multi-cloud or hybrid settings. Finally, Graylog lacks the real-time efficiency of ELK or Fluentd and is less optimized for edge deployments due to its resource demands.

\mypar{Continuous Integration and Deployment} 
Continuous Integration (CI) and Continuous Deployment (CD) are software engineering practices that enhance development efficiency and software quality. CI involves frequent integration of code changes into a central repository, enabling automated testing and faster bug detection. CD complements CI by automating the build, test, and deployment processes, ensuring software is always in a releasable state. Frameworks like \textit{ArgoCD}\footnote{\url{https://argo-cd.readthedocs.io/}},
and \textit{Flux}\footnote{\url{https://fluxcd.io/}}, built for \textit{Kubernetes}, provide declarative, Git-based pipelines that enable versioned and automated deployments across edge clusters and cloud systems.
Spinnaker\footnote{\url{https://spinnaker.io/}} is a powerful multi-cloud deployment orchestration tool with robust support for edge deployments, enabling organizations to manage complex deployment pipelines with canary releases, blue-green deployments, and rolling updates across multiple cloud providers.
Jenkins\footnote{\url{https://www.jenkins.io/}} is a widely used CI/CD tool that supports a variety of integrations, including \textit{Kubernetes} for container orchestration, \textit{Terraform} for infrastructure as code (IaC), \textit{GitHub Actions} and \textit{GitLab CI} for source code management, as well as cloud services like \textit{AWS CodeBuild}, \textit{Azure DevOps}, and \textit{Google Cloud Build} for scalable deployment automation. Finally, Tekton\footnote{\url{https://tekton.dev/}} is a lightweight, stateless, and cloud-native CI/CD framework specifically designed for Kubernetes-native pipelines.
Major cloud providers also offer native CI/CD services integrated into their ecosystems. For instance, \textit{AWS CodePipeline} automates release pipelines, integrating seamlessly with AWS services like \textit{AWS CodeBuild} and \textit{AWS CodeDeploy}. Similarly, \textit{Azure Pipelines} supports multi-platform builds and deployments with strong integration with Kubernetes. \textit{Google Cloud Build} provides a serverless platform for automating builds, tests, and deployments across hybrid environments, including on-premises, multi-cloud, and hybrid cloud setups.

\mypar{Monitoring} Monitoring applications and systems in the edge-cloud continuum requires real-time insights into performance across geographically distributed environments. Open-source tools like Prometheus\footnote{\url{https://prometheus.io/}} support time-series metrics collection for multi-cluster setups, providing a scalable solution for distributed monitoring. 
Grafana\footnote{\url{https://grafana.com/}} complements Prometheus by offering powerful data visualization capabilities. It allows users to create customizable dashboards that seamlessly integrate with Prometheus and other data sources, enabling unified insights into system health and performance.
Thanos\footnote{\url{https://thanos.io/}} extends Prometheus by enabling global querying and long-term storage, making it ideal for hybrid monitoring across edge and cloud systems. It is designed to operate in multi-cluster setups and supports highly distributed architectures, ensuring visibility across both edge nodes and cloud infrastructures.
Commercial solutions such as Datadog\footnote{\url{https://www.datadoghq.com/}} and New Relic\footnote{\url{https://newrelic.com/}} offer full-stack observability tailored to the edge-cloud continuum. These tools provide advanced features such as distributed tracing, log correlation, and application performance monitoring, making them highly effective for heterogeneous systems. Datadog, for instance, integrates seamlessly with IoT devices, servers, and cloud platforms. New Relic, instead, focuses on providing a unified view of application and infrastructure performance, offering AI-driven insights to optimize operations in hybrid environments. 
Cloud providers also offer robust monitoring solutions for the edge-cloud continuum, which integrate closely with their cloud ecosystems. For instance, \textit{Azure Monitor} delivers end-to-end observability for Azure resources, on-premises systems, and hybrid environments, combining metrics, logs, and traces in a unified platform. \textit{AWS CloudWatch} provides similar capabilities for Amazon Web Services, enabling users to monitor applications, services, and IoT devices while offering alerting and automated actions to ensure system health. \textit{Google Cloud Operations Suite} integrates monitoring, logging, and diagnostics for Google Cloud environments, with features like real-time alerts and root cause analysis for issue resolution. 

\section{Research Outlook}
\label{sec:challenges}

As outlined in previous sections, achieving a seamless edge-cloud continuum requires addressing key challenges across \textit{infrastructure}, \textit{services}, and \textit{applications}. This section examines the main open issues and potential solutions, along with emerging trends that are shaping the future of these systems.

\subsection{Open Challenges}
\mypar{Heterogeneity and Interoperability}
A significant hurdle in the edge-cloud continuum is the vast heterogeneity of devices and platforms involved, each with varying computational capabilities, architectures, and communication protocols~\cite{gkonis2023survey}. Managing and ensuring seamless interoperability among these diverse components is considerably more complex than in homogeneous cloud environments, also due to the lack of standardized connection and programming protocols. Addressing this requires the development of hardware and technology-agnostic protocols, along with the adoption of open-source frameworks and middleware platforms to facilitate integration across diverse environments~\cite{moreschini2022cloud}. Future trends point towards greater emphasis on standardization efforts and the development of unified programming abstractions to simplify the utilization of these heterogeneous resources.

\mypar{Resource Management and Orchestration}
Efficiently managing and orchestrating computational, storage, and network resources across geographically distributed nodes presents a complex optimization problem~\cite{soumplis2022resource}. This involves dynamically offloading tasks between edge devices, intermediate far- and near-edge nodes, and the cloud based on application requirements, resource availability, network conditions, and energy constraints. Existing orchestration tools, primarily designed for cloud-based deployments, often fall short in addressing the dynamic characteristics of edge environments, such as device mobility and fluctuating network conditions, which instead necessitate resource management systems capable of real-time adaptation~\cite{soumplis2022resource}. Overcoming this challenge requires the development of intelligent task offloading algorithms and optimal resource allocation mechanisms tailored for heterogeneous edge-cloud environments, which machine learning offering a promising approach for achieving near-optimal solutions in these complex scenarios~\cite{gkonis2023survey}.

\mypar{Security and Privacy}
Ensuring the security and privacy of data and applications across distributed environments is crucial due to the expanded attack surface and the presence of sensitive data at the network's edge~\cite{arzovs2024distributed}. The transfer of data between edge devices and the cloud necessitates robust security and privacy enhancements to counter various threats~\cite{gkonis2023survey}. The inherent distribution of the continuum, often involving devices owned by different entities, introduces complexities in establishing trust and overall reliability. Moreover, device heterogeneity implies varying levels of security capabilities, with resource-constrained IoT devices potentially lacking the capacity for complex encryption. Addressing these concerns requires implementing secure access mechanisms, robust encryption protocols, and effective authentication methods across the continuum. Privacy-preserving techniques like federated learning are also crucial for enabling distributed learning while safeguarding sensitive data residing on edge devices~\cite{kumar2020federated}.

\mypar{Energy Efficiency and Sustainability}
Edge devices often operate with limited battery power, and the collective energy consumption of a large number of distributed edge nodes can be substantial~\cite{krekovic2025reducing}. Ensuring energy efficiency and promoting sustainability are therefore critical considerations, particularly for large-scale deployments of the edge-cloud continuum. Processing data at the edge can reduce the amount of data transmitted to the cloud, leading to conservation of network bandwidth and energy. However, it is equally important to minimize the energy footprint of the edge devices themselves. Optimal resource allocation strategies should explicitly consider the energy consumption of the nodes involved in processing and communication. Addressing this challenge necessitates the development of communication protocols designed to minimize energy consumption and the implementation of energy-aware task scheduling algorithms.

\mypar{Data Management and Distributed Analytics}
Managing the ever-increasing volume, velocity, and variety of edge-generated data and ensuring consistency across the diverse components of the continuum is a significant challenge~\cite{gkonis2023survey}. Distributed data management involves tasks such as data collection, aggregation, filtering, and ensuring transparent data access. Maintaining data consistency between the edge, where initial processing often occurs, and the cloud, which serves as a long-term storage and analytics repository, is crucial for data integrity. The need to process data close to its source for low-latency applications necessitates intelligent data management strategies to determine optimal processing locations~\cite{krekovic2025reducing}. Addressing this challenge involves developing efficient techniques for data collection, filtering, and pre-processing at the edge, along with effective data synchronization mechanisms across distributed nodes.

\mypar{Standardization and Policy Frameworks}
The edge-cloud continuum is a relatively new computing paradigm, and consequently, a comprehensive set of standards and mature development frameworks to guide its implementation and widespread adoption are still lacking. While various standards organizations and open-source communities are actively working on defining specifications for specific aspects, a holistic and unified set of standards encompassing the entire continuum is still evolving. This absence of well-established development frameworks can significantly hinder the process of application development and deployment across the edge-cloud continuum. Addressing this challenge requires the development of user-friendly and comprehensive development tools and frameworks to facilitate application creation and deployment~\cite{rodrigues2022exploiting, belcastro2024developing}.

\subsection{Future Trends}
Looking ahead, the evolution of edge-cloud continuum systems is expected to be strongly influenced by several key trends, particularly related to the rapid advancement of AI technologies. The integration of advanced AI models, including generative AI, into the edge-cloud continuum presents unique challenges due to their substantial computational requirements, memory footprint, and energy consumption.
\textit{Generative AI} tools and models, such as LLMs, can enhance developer productivity within the edge-cloud continuum by automating code generation, providing intelligent recommendations, early identification of potential security issues, and facilitating debugging and maintenance tasks~\cite{DEVITO2025107829}. 
However, a major challenge is achieving a comprehensive understanding and holistic view of the entire application architecture, codebase, and associated components. Generative AI systems must accurately interpret complex architectural dependencies and interactions to effectively detect errors, anticipate compatibility issues, and suggest contextually relevant solutions. Additionally, ensuring the correctness, efficiency, and security of AI-generated code across diverse hardware and software environments necessitates robust validation and testing frameworks. Future developments should focus on advanced generative models capable of contextual awareness, architectural interpretation, and integration within validation mechanisms to ensuring reliable, secure, and efficient AI-assisted programming. 
Moreover, \textit{AI Agent} systems represent a promising trend within the edge-cloud continuum by providing autonomous decision-making, intelligent task execution, and effective coordination across distributed components. These AI agents can independently assess their environment, communicate with other agents, and dynamically adapt their behavior based on changing conditions. Challenges in deploying agent-based systems include managing decentralized control, ensuring seamless communication between diverse agents, and maintaining robust performance in highly dynamic and resource-constrained environments. Future directions involve developing standardized communication protocols, advanced negotiation algorithms, and reliable mechanisms for coordination and conflict resolution among agents.
Finally, integrating edge computing with interactive robotic systems, including humanoids, poses unique challenges due to the requirement for seamless, real-time \textit{human-robot interactions} (HRI). Edge devices such as robots must interpret human gestures, speech, emotions, and contextual cues reliably to provide meaningful interactions. Achieving this demands significant computational capabilities, advanced sensing technologies, and sophisticated AI algorithms optimized for low latency and high accuracy. Furthermore, ensuring user safety, trustworthiness, and adaptability in highly dynamic and unpredictable human environments complicates these deployments. Future research will focus on developing robust interaction frameworks, optimizing real-time communication protocols, and enhancing AI models capable of nuanced human understanding and adaptive responsiveness in resource-constrained edge settings.

\section{Conclusion}
\label{sec:conclusion}

The edge-cloud continuum represents a fundamental shift in the design and deployment of distributed computing systems, addressing the growing demand for low-latency, privacy-preserving, and scalable data processing. However, realizing the full potential of this paradigm remains challenging due to infrastructural disparities, fragmented standards, and the complexity of orchestrating services across heterogeneous environments. 
This survey tackles these challenges by offering a comprehensive, developer-centric perspective on the edge-cloud continuum. Through a structured framework, we bridge theoretical foundations with practical insights, delivering a state-of-the-practice survey that examines architectural models, computational paradigms, enabling technologies, and deployment platforms. We also highlight real-world application domains and provide an overview of testing tools and benchmarking strategies essential for effective implementation. By integrating insights from both academic research and industry developments, this work serves as both a practical guide for developers and a foundational reference for researchers. Our analysis of public and private platform capabilities, alongside an exploration of key service orchestration strategies, is intended to inform best practices and guide strategic decisions in the design and deployment of edge-cloud systems. Finally, we describe several open challenges that continue to impact this field, including the need for standardized interfaces, adaptive resource management strategies, and globally distributed infrastructure to ensure equitable access and consistent performance. We also discuss key future trends, particularly those related to emerging AI developments, which are expected to further influence the evolution and capabilities of edge-cloud systems.

\bibliographystyle{ACM-Reference-Format}
\bibliography{references}

\begin{acks}
This work was supported by the research project “INSIDER: INtelligent ServIce Deployment for advanced cloud-Edge integRation” granted by the Italian Ministry of University and Research (MUR) within the PRIN 2022 program and European Union - Next Generation EU (grant n. 2022WWSCRR,  CUP H53D23003670006) and by the European Union under the Italian National Recovery and Resilience Plan (NRRP) of NextGenerationEU, partnership on “Telecommunications of the Future” (PE00000001 - program “RESTART”). We also acknowledge support by the ``National Centre for HPC, Big Data and Quantum Computing'' project, CN00000013 - CUP H23C22000360005.
\end{acks}

\end{document}